\documentclass[apj]{emulateapj}

\usepackage{lscape}

\slugcomment{ApJS, in press}

\shorttitle{X-Ray Variability in NGC 6946, NGC 4485/90}
\shortauthors{Fridriksson et al.}

\begin{document}

\title{The Long-Term Variability of the X-Ray Sources in NGC 6946 and NGC 4485/4490}

\author{Joel K.~Fridriksson\altaffilmark{1,2}, Jeroen Homan\altaffilmark{2}, Walter H.~G.~Lewin\altaffilmark{1,2}, Albert K.~H.~Kong\altaffilmark{3}, and David Pooley\altaffilmark{4}}

\altaffiltext{1}{Department of Physics, Massachusetts Institute of Technology, Cambridge, MA 02139; joelkf@mit.edu.}
\altaffiltext{2}{MIT Kavli Institute for Astrophysics and Space Research, Cambridge, MA 02139.}
\altaffiltext{3}{Institute of Astronomy and Department of Physics, National Tsing Hua University, Hsinchu, Taiwan 30013.}
\altaffiltext{4}{Department of Astronomy, University of Wisconsin--Madison, Madison, WI 53706.}

\begin{abstract}
We analyze data from five {\it Chandra} observations of the spiral galaxy NGC~6946 and from three {\it Chandra} observations of the irregular/spiral interacting galaxy pair NGC~4485/4490, with an emphasis on investigating the long-term variability exhibited by the source populations. We detect 90 point sources coincident with NGC~6946 down to luminosities of a few times $10^{36}\textrm{ ergs s}^{-1}$, and 38 sources coincident with NGC~4485/90 down to a luminosity of $\sim1\times10^{37}\textrm{ ergs s}^{-1}$. Twenty-five (15) sources in NGC~6946 (NGC~4485/90) exhibit long-term (i.e., weeks to years) variability in luminosity; 11 (4) are transient candidates. The single ultraluminous X-ray source (ULX) in NGC~6946 and all but one of the eight ULXs in NGC~4485/90 exhibit long-term flux variability. Two of the ULXs in NGC~4485/90 have not been identified before as ultraluminous sources. The widespread variability in both systems is indicative of the populations being dominated by X-ray binaries, and this is supported by the X-ray colors of the sources. The distribution of colors among the sources indicates a large fraction of high-mass X-ray binaries in both systems. The shapes of the X-ray luminosity functions of the galaxies do not change significantly between observations and can be described by power laws with cumulative slopes $\sim0.6-0.7$ (NGC~6946) and $\sim0.4$ (NGC~4485/90).
\end{abstract}


\section{Introduction}\label{sec:intro}
The systematic study of X-ray source populations in external galaxies first became possible with the {\it Einstein} Observatory in 1978. Significant advances in the field were made with subsequent satellites such as {\it ROSAT}, but a giant leap forward has been taken with {\it Chandra}. With its subarcsecond spatial resolution and high sensitivity it is possible to resolve the vast majority of the luminous ($\gtrsim10^{37}\textrm{ ergs s}^{-1}$) X-ray sources in galaxies out to distances of $20-30$ Mpc. In addition, the spectrometric capabilities of the {\it Chandra} ACIS CCD detector allow spectral properties of sources to be extracted. For two recent reviews of the study of X-ray sources in normal galaxies, emphasizing results from {\it Chandra}, see~\cite{fabbiano2006a} and~\cite{fabbiano2006b}.

We now know that X-ray source populations in galaxies are dominated at high luminosities ($\gtrsim10^{37}\textrm{ ergs s}^{-1}$, the range typically explored with {\it Chandra} observations) by X-ray binaries (XRBs) consisting of an accreting neutron star or black hole and a stellar companion. In addition, galaxies usually have a few young supernova remnants (SNRs) in this luminosity range. Unsurprisingly, the X-ray populations in early-type galaxies (E and S0) seem to consist mostly of low-mass X-ray binaries (LMXBs), whereas galaxies with younger stellar populations (spiral and irregular galaxies) typically have a much higher fraction of the shorter lived high-mass X-ray binaries (HMXBs). In galaxies with a high star formation rate HMXBs are especially common.

An important class of sources in external galaxies are the so-called ultraluminous X-ray sources (ULXs), usually defined as non-nuclear sources with implied isotropic X-ray luminosities $\geq10^{39}\textrm{ ergs s}^{-1}$. Detections of ULXs with luminosities as high as $\sim10^{41}\textrm{ ergs s}^{-1}$ have been reported \citep[see, e.g.,][]{davis2004}. The nature of ULXs is still debated, and it has been argued that at least some of them might be a new class of objects, so-called intermediate-mass black holes (IMBHs) with masses $M\sim10^2-10^4\textrm{ }M_{\sun}$ \citep[see, e.g.,][]{miller2004}.

We present in this paper the X-ray source population study of the spiral galaxy NGC 6946 and the interacting irregular/spiral system NGC 4485/4490. A special emphasis is placed on studying the long-term (weeks to years) variability properties of the source populations. These galaxies were chosen because they are nearby ($D\lesssim8$ Mpc) and have multiple (three or more) long ($\gtrsim 20$ ks) {\it Chandra} ACIS exposures spanning a baseline of a few years. Both show an enhanced star formation rate. The spiral NGC 6946 also has the fortunate characteristic of being observed nearly face-on, and the NGC 4485/90 system has low Galactic extinction and a large number of ULXs. For more background information on the galaxies, see \S\S~\ref{sec:ngc6946} and \ref{sec:ngc4485}.

Not much work has been done on the long-term variability of X-ray sources in external galaxies, since {\it Chandra} observations at multiple epochs are usually not available. We know from observations in our own Galaxy that time variability of various kinds, including transient outbursts, eclipses, dips, as well as less severe variations in flux, is very common among XRBs. Temporal and spectral analysis carried out for the most luminous sources in nearby galaxies shows behavior similar to that in Galactic XRBs, clearly pointing to XRB populations \citep[see][and references therein]{fabbiano2006b}. For example, \cite{kong2002} find that among 204 detected sources in M31, 50\% are variable on timescales of months and 13 are transients. \citet{sivakoff2005b} find short-timescale flares in 3 out of 157 sources in the elliptical galaxy NGC 4697, and two of the flares have durations and luminosities similar to Galactic superbursts \citep[thermonuclear bursts with very long (hours) durations and very large fluences; see][]{kuulkers2004}. \cite{sivakoff2005a} also find long-term variability in 26 out of 124 sources in NGC 4697, and 11 of those are transient candidates. \cite{zezas2006}, analyzing seven {\it Chandra} observations of the Antennae galaxies, find intensity and/or spectral variability among sources on timescales of years, months, days, and hours. Overall, $\sim50\%$ of the sources detected in each observation show either spectral or intensity variation but do not all follow a common trend, indicating that there are various classes of sources. Of the 14 ULXs in the Antennae, 12 show long-term variability. In general, variability of some sort is very common among ULXs \citep[see][]{fabbiano2006a, fabbiano2006b}. Despite widespread variability in luminosity among individual sources, X-ray luminosity functions (XLFs) seem to be remarkably stable from one observation to another, as indicated by observations of NGC 5128 \citep{kraft2001}, M33 \citep{grimm2005}, and the Antennae \citep{zezas2007}.

The organization of the paper is as follows. In \S~\ref{sec:analysis} we describe the common analysis steps performed for both galaxies, including source detection, photometry, the construction of light curves and hardness ratios, and testing for flux and spectral variability. In \S~\ref{sec:observations} we discuss general properties and the observations of the galaxies and present the results of the source detection. In \S~\ref{sec:properties} we present and discuss our results on various properties of the source populations in the galaxies. Finally, in \S~\ref{sec:conclusions} we briefly summarize our results. Included are tables with the source catalog for each galaxy and various observed and derived parameters for each source.

\section{Data Analysis}\label{sec:analysis}
The {\it Chandra} observations of both galaxies were analyzed with the CIAO software, version 3.4 (CALDB ver. 3.3.0), and with ACIS Extract (hereafter AE), version 3.131. AE is an ACIS point source extraction package developed at Pennsylvania State University \citep{broos2002}, which assists in a large variety of complex data processing tasks. The procedures used in AE are described in \citet{TFM03} and \citet{Getman05}. We note that all errors mentioned in this paper are, unless explicitly stated otherwise, Poisson errors with a confidence level corresponding to 1 $\sigma$ Gaussian. They were computed using the Gehrels approximation of Poisson limits \citep{gehrels1986}.

\subsection{Source detection}\label{sec:detection}

The level 1 event files from the {\it Chandra} data archive were reprocessed, and new level 2 event files created by following the standard ACIS data preparation procedure recommended by the {\it Chandra} X-ray Center (CXC).\footnote{See \url{http://cxc.harvard.edu/ciao/guides/acis\_data.html}.} The data were checked for periods of unusually high background. No such periods were found in any observation of either galaxy. A monochromatic exposure map evaluated at 1.5 keV was created for each observation, and the event files, as well as the exposure maps, merged to create a co-added exposure covering all the observations. A $0.3-7$ keV band image encompassing the whole galaxy was subsequently created from the merged event file, and \verb#wavdetect#, the CIAO wavelet source detection algorithm, run at nine scales (1, $\sqrt{2}$, 2, $2\sqrt{2}$, 4, $4\sqrt{2}$, 8, $8\sqrt{2}$, and 16 pixels), using the merged exposure map (the \verb#sigthresh# parameter, which sets the threshold for identifying a pixel as belonging to a source, was set at a value of 1 over the number of pixels in the image\footnote{See \url{http://cxc.harvard.edu/ciao/ahelp/wavdetect.html}.}).

Further analysis was performed with AE. The input to the program was the master source list from the \verb#wavdetect# run and the cleaned event data for each observation. After visual inspection of the data (both co-added and individual exposures) a few extra candidate sources, not detected by \verb#wavdetect#, were typically added to the list. Since we only perform source detection on the co-added exposure, it is conceivable that \verb#wavdetect# might for example miss weak transient sources in high-background regions. In the end, only in one case did such a candidate source end up surviving the significance cuts described below. Source CXOU J123038.2+413831 in NGC 4485/90 is in close proximity to a much brighter transient source, causing \verb#wavdetect# to detect only the brighter source in the co-added exposure. The existence of two individual sources is readily apparent when looking at individual observations, in some of which the brighter source is in quiescence.

An important characteristic of AE is that it tailors the count extraction region individually to each source in each observation based on the shape of the point spread function (PSF) at the source location on the detector. Counts were by default extracted from a region with a PSF encircled energy fraction of 0.90, but in cases of crowded sources AE automatically reduces the fraction to prevent overlapping extraction regions. The response functions computed by AE are corrected to take into account the flux missed by the finite extraction region in each case.

Possibly spurious detections from our candidate source list were rejected based on a source significance statistic computed by AE \citep[for definition see][]{broos2002}, along with visual inspection of the individual and merged observations, as well as the \verb#wavdetect# significance statistic.\footnote{See definition in The {\it Detect} Reference Manual, available online at \url{http://cxc.harvard.edu/ciao/download/doc/detect\_manual/index.html}.} We set a conservative limit where detections with AE significance $\lesssim3-3.5$ $\sigma$ (which typically corresponds to \verb#wavdetect# significance $\lesssim5-6$ $\sigma$ for the co-added exposure) were rejected.

The source positions were overlaid on an optical image of the galaxy along with the $D_{25}$ elliptical isophote \citep[obtained from][]{rc3}. In general, we considered sources inside the $D_{25}$ isophote to be coincident with the system and removed other sources from the list. However, in a few cases sources just outside the isophote were included based on visual inspection of the galaxy's optical extent, which can be quite irregular. The accuracy of the source positions was refined with the help of AE. Since the average off-axis angle was $\lesssim5'$ for nearly all sources, we used positions based on the centroid of the extraction regions as recommended in \cite{broos2002}.

\subsection{Light Curves}\label{sec:light_curves}

AE is capable of performing extensive broadband photometry. For all sources an estimate of both the photon flux and the energy flux in the $0.3-7$ keV band was computed from the AE results for each observation, as well as for the co-added exposure. For all but the faintest sources the photon flux was computed by dividing the number of net (i.e., background-subtracted) counts by the exposure time and the mean effective area at the source location for a given energy band (the MEAN\_ARF parameter in AE, which takes into account flux missed by the finite extraction region). For computing the mean effective area the simplifying assumption of a flat incident spectrum within the given energy band was made. For an estimate of the energy flux the photon flux was multiplied by the mean observed photon energy in the corresponding band, as computed by AE. This was done in five adjacent energy bands spanning together the full $0.3-7$ keV band ($0.3-1$, $1-1.5$, $1.5-2.5$, $2.5-4$, and $4-7$ keV), and the results from the individual bands then added together. This is a compromise between calculating the fluxes directly for the full band (and thus assuming a flat incident spectrum over that whole band, which gives rise to a systematic error) and dividing the counts in each spectral channel by the corresponding value of the auxiliary response function (ARF), this latter method being susceptible to large Poisson errors. For the ULXs in the galaxies we performed spectral fitting (see below) and compared the flux results from our ``band method'' to the spectral fitting results. We found that the values from the band method were typically $5\%-10\%$ higher than the ones from spectral fitting and that the discrepancy was usually slightly greater for the photon flux than the energy flux.

For observations of sources with very few counts it is perhaps more sensible to give an upper limit to the flux instead of an actual flux estimate. Since we do not perform source detection on individual observations, only on the co-added exposure, it is effectively meaningless for us to talk about whether a source was detected or not in an observation. We therefore adopt a fixed threshold of five net counts, below which we give upper limits in our tables, with 3 $\sigma$ Gaussian confidence, instead of flux values. We calculate the limits with the Sherpa package in CIAO. They are derived from the full-band counts, assuming a power-law spectrum with a photon index of 1.7 (characteristic of best-fit spectra for sources in our sample) and Galactic absorption, and using the response functions at the source location. We note that for the purpose of computing the flux variability parameter described in \S~\ref{sec:variability} below, we always use actual flux estimates, although for less than 5 net counts those estimates are calculated using Sherpa (as described above) rather than our band method.

In the case of the ULXs the flux estimates were improved on by fitting eight simple spectral models to each of the {\it Chandra} observations in XSPEC, and using in each case the best-fitting model to derive the fluxes. The models considered were absorbed power-law, bremsstrahlung, blackbody, and multicolor disk blackbody models, as well as two-component models with one of the previous components plus an additional power-law component (both equally absorbed). The absorption parameter was allowed to vary freely with a minimum value equal to the Galactic absorption column.

After converting the energy fluxes to luminosities based on our adopted distance to each galaxy long-term light curves (spanning all observations) were created for each source. It should be noted that all fluxes and luminosities quoted in this paper are absorbed values (i.e., they are not corrected for absorption along the line of sight), and they are, unless otherwise noted, given for the $0.3-7$ keV range. The error bars shown in light curves and other plots only take into account the Poisson errors in the counts.

\begin{deluxetable}{lcccc}
\tablewidth{0pt}
\tabletypesize{\scriptsize}
\tablecaption{{\it Chandra} Observations of NGC 6946\label{obs6946chandra}}
\tablehead{\colhead{Start Date} & \colhead{Instrument} & \colhead{Obs. ID} & \colhead{Exp.\tablenotemark{a} (ks)} & \colhead{Data Mode\tablenotemark{b}}}
\startdata
2001 Sep 7  & ACIS-S & 1043 & 58.29 & F \\
2002 Nov 25 & ACIS-S & 4404 & 29.92 & F \\
2004 Oct 22 & ACIS-S & 4631 & 29.70 & F \\
2004 Nov 6 & ACIS-S & 4632 & 27.94 & F \\
2004 Dec 3 & ACIS-S & 4633 & 26.62 & F \\
\enddata
\tablenotetext{a}{Good live exposure time.}
\tablenotetext{b}{Telemetry formats: $\textrm{F}=\textrm{Faint}$.}
\end{deluxetable}

\subsection{Hardness ratios}\label{sec:hardness_ratios}

Hardness ratios were chosen as our primary means of investigating the spectral properties of the source populations, since a large portion of the sources had too few counts to be amenable to spectral fitting. Two kinds of hardness ratios were computed for each source in each observation, as well as for the co-added exposure. Our soft color is defined as
\begin{equation}
\textrm{HR1}=\frac{\textrm{M}-\textrm{S}}{\textrm{T}},
\end{equation}
and the hard color as
\begin{equation}
\textrm{HR2}=\frac{\textrm{H}-\textrm{M}}{\textrm{T}},
\end{equation}
where S, M, H, and T are the net counts in the $0.3-1$ keV (soft), $1-2$ keV (medium), $2-7$ keV (hard), and $0.3-7$ keV (total) bands, respectively. The counts used to calculate the hardness ratios were in each case divided by the mean value of the effective area over the given energy band at the location of the source. This was done to account for differences in effective areas between different energy bands, and differences due to variations in source locations on the detector. The errors in the hardness ratios were computed using standard error propagation.

Although it is not possible to conclusively classify an individual source based on its hardness ratios alone, it is possible to draw conclusions about bulk trends within a source population, as shown by \citet{prestwich2003}. They propose a classification scheme based on hardness ratios, in which they identify the regions in a color-color diagram where sources of a given type are most likely to fall. We utilize this method to draw conclusions about the source populations (see \S~\ref{sec:spec_class}). It should be noted that our hardness ratio definitions differ slightly from the ones used by \citet{prestwich2003}. They use an upper limit of 8 keV to their hard and total bands, in contrast to our using 7 keV. However, for the vast majority of sources only a very small fraction of the counts in the $0.3-8$ keV band is between 7 and 8 keV, so this should have a negligible effect. They also use a slightly different method to account for variations in effective area. As \citet{prestwich2003} discuss in their paper, absorption will tend to blur the distinction in a color-color diagram between the different source types. However, this effect should not be very severe for the galaxies studied in this paper; although NGC 6946 suffers from substantial extinction, it is observed face-on, and the absorption column in the direction of NGC 4485/90 is low (see \S~\ref{sec:observations}).

\subsection{Variability}\label{sec:variability}

We compute a significance parameter for long-term flux variability for each source:
\begin{equation}
S_\mathrm{flux}=\mathrm{max}_{i,j}\frac{|F_i-F_j|}{\sqrt{\sigma_{F_i}^2+\sigma_{F_j}^2}},
\end{equation}
where $F_i$ denotes the photon flux of a source in the $i\textrm{th}$ {\it Chandra} observation, and $\sigma_{F_i}$ is the corresponding error. We define a source as variable if $S_\mathrm{flux}>3$. We also quantify the variability by computing the ratio $R=F_\mathrm{max}/F_\mathrm{min}$. We consider a source to be a transient candidate if its measured flux is consistent with zero (within $\sim1$ $\sigma$ error) in at least one observation, and if the source is also clearly variable ($S_\mathrm{flux}>3$). 

We test for short-term flux variability (within single observations) by performing a one-sided Kolmogorov-Smirnov test for each source in each observation, comparing event times to a uniform count-rate model. We define a source as showing evidence for variability if the significance of the computed K-S statistic is less than $3\times10^{-3}$ (i.e., if the null hypothesis that the count rate is uniform can be rejected at the 99.7\% level or more, corresponding to approximately 3 $\sigma$ Gaussian confidence).

To search for long-term spectral variability, we compute significance parameters for the variability in the hardness ratios. They are defined analogously to the flux variability parameter above. Again, we require 3 $\sigma$ confidence to classify a source as variable.

\subsection{{\it ROSAT} and {\it XMM-Newton} Data}\label{sec:rosat_xmm}

Apart from {\it Chandra}, there are three other X-ray telescopes that have had good enough spatial resolution to possibly be of significant value to this study. The {\it Einstein} HRI, the {\it ROSAT} HRI, and the {\it XMM-Newton} EPIC cameras all had or have an on-axis resolution of $\lesssim10''$. However, the {\it Einstein} and {\it ROSAT} HRIs were only sensitive in the $0.15-3.5$ and $0.1-2.4$ keV bands and had negligible energy resolution. None of the galaxies analyzed in this paper was observed with the {\it Einstein} HRI.

For the ULXs that were well resolved from neighboring bright sources the long-term light curves were extended by adding data points from {\it ROSAT} and {\it XMM-Newton} observations. It was usually necessary to use small extraction regions to minimize the contamination from nearby sources and from the diffuse emission in which many sources are embedded.

\begin{deluxetable}{lcc}
\tablewidth{0pt}
\tabletypesize{\scriptsize}
\tablecaption{{\it ROSAT} HRI Observations of NGC 6946\label{obs6946rosat}}
\tablehead{\colhead{Start Date} & \colhead{Seq. ID} & \colhead{Exp.\tablenotemark{a} (ks)}}
\startdata
1994 May 14 & rh600501n00 & 59.89\\
1995 Aug 13 & rh600718n00 & 21.51\\
1996 Aug 10 & rh500476n00 & 8.36\\
\enddata
\tablenotetext{a}{Good live exposure time.}
\end{deluxetable}

\subsubsection{{\it ROSAT}}
{\it ROSAT} count rates were derived from the RDF (rationalized data format) files available from the {\it ROSAT} Archive. The event data files were corrected for the HRI aspect error \citep[see, e.g.,][]{david1999} using the correction script developed at the Smithsonian Astrophysical Observatory (SAO).\footnote{Available online at \url{http://hea-www.harvard.edu/rosat/aspfix.html}.} Background-subtracted count rates were extracted using the ASTERIX software package (ver. 2.3-b2). The \verb#xrtsort#, \verb#xrtsub#, and \verb#xrtcorr# tasks were used, which automatically make dead-time, vignetting, and quantum efficiency corrections. In all cases counts were extracted from a circle of radius $8\arcsec$. For point sources with an off-axis angle $\lesssim5'$ the PSF encircled energy fraction for such a circle is $\sim0.84$ \citep{david1999, zimmermann1998}. The count rates were therefore divided by this number to correct for the counts missed by the finite extraction region.

To facilitate comparison with the {\it Chandra} results count rates from {\it ROSAT} were converted to energy flux in the $0.3-7$ keV band using PIMMS. To determine the spectral model to use for the conversion the eight models described in \S~\ref{sec:light_curves} were fitted to the source spectrum from the co-added {\it Chandra} exposure, and the best-fitting model used.

\begin{figure*}
\centering
\includegraphics[height=8.5cm]{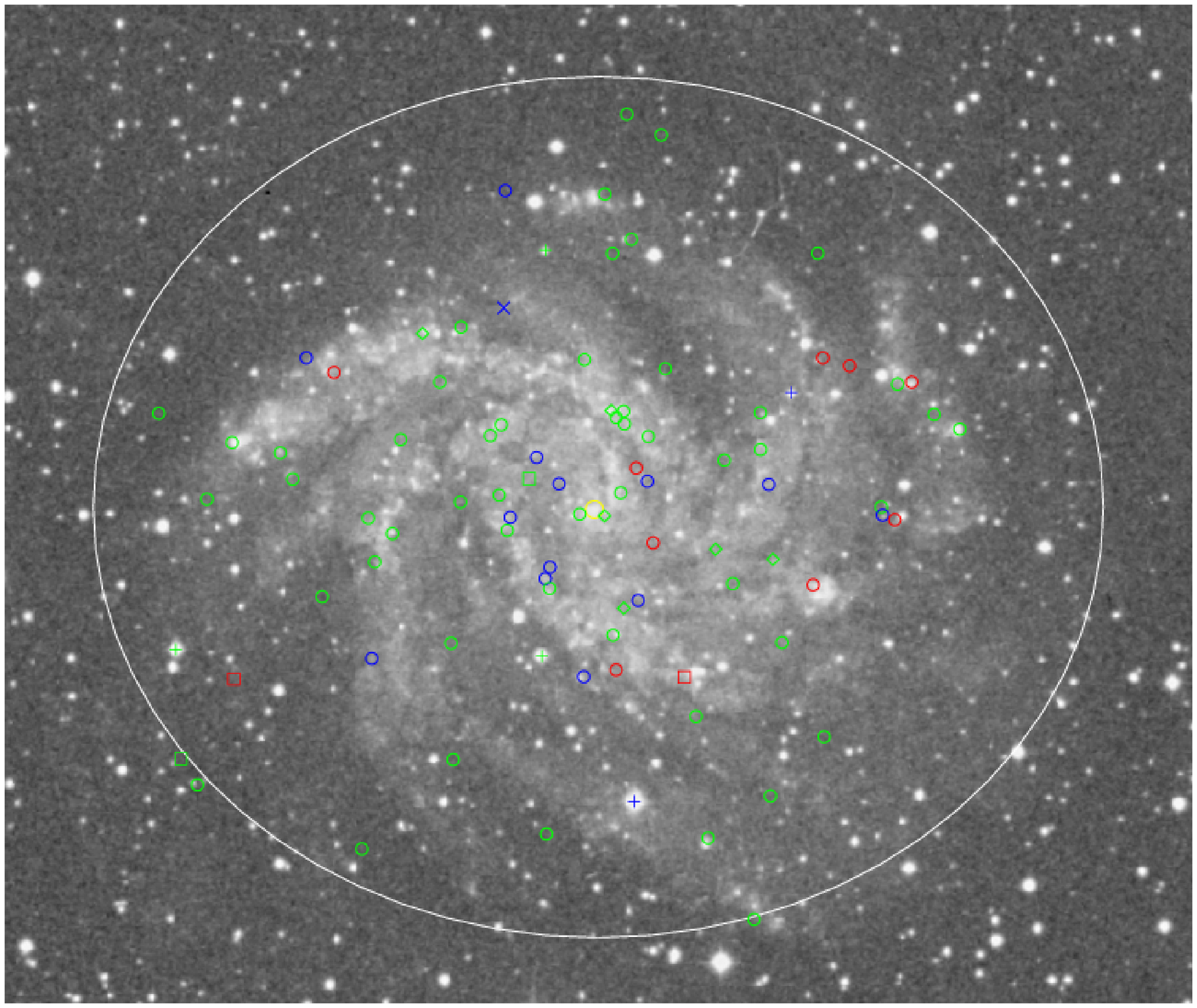}
\phantom{}
\includegraphics[height=8.5cm]{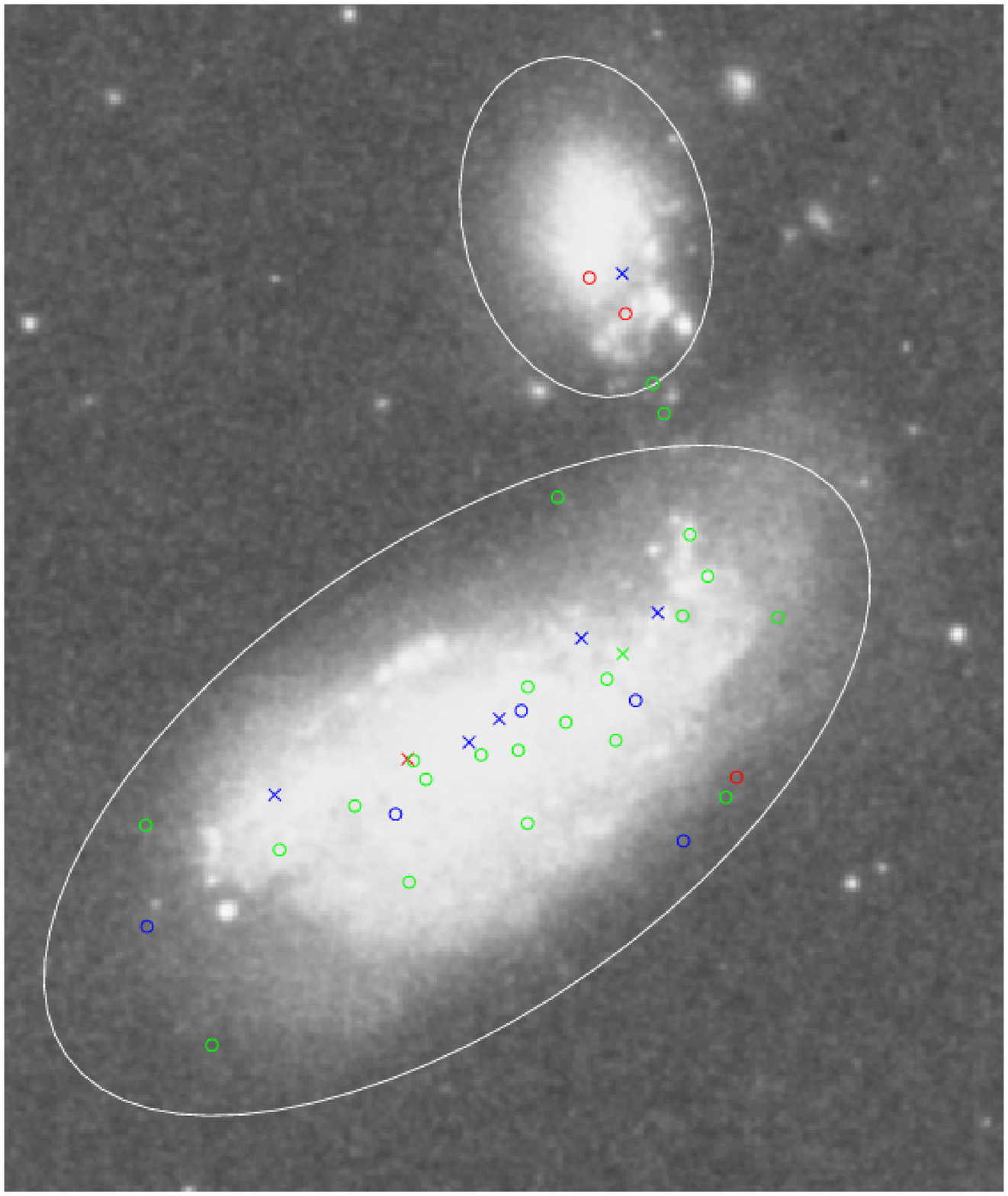}
\caption{DSS images of NGC 6946 ({\it left}) and NGC 4485/4490 ({\it right}) with positions of detected X-ray sources overlaid. Foreground stars are represented with plus symbols, historical supernovae with boxes, SNRs with diamonds, ULXs with crosses, and other sources with circles. Blue indicates that long-term variability is detected, green that it is not, and red denotes transient candidates. The white ellipses are the $D_{25}$ elliptical isophotes and the yellow circle encloses the central area in NGC 6946 that was excluded from analysis.}\label{fig:dss}
\end{figure*}

\subsubsection{{\it XMM-Newton}}
For {\it XMM-Newton} data a new analysis was performed for each of three EPIC detectors using the SAS software, versions 7.0.0 and 7.1.0, and the latest available calibration files at the time of the analysis. Starting with the original data files (ODFs) from the {\it XMM-Newton} archive, MOS and PN data were reprocessed using the \verb#emproc# and \verb#epproc# tasks. After excluding periods of high background, source spectra were extracted with the \verb#evselect# task. The \verb#eregionanalyse# task was used to center the extraction radius on the centroid of the counts distribution for each source. Counts were extracted from circles with radii in the range $8''$ to $15''$ as deemed appropriate in each case. The event selection criteria used for the spectral extraction were $\textrm{PATTERN}=0-12$ and $\textrm{FLAG}=\#\textrm{XMMEA\_EM}$ for the MOS instruments, and $\textrm{PATTERN}=0-4$ and $\textrm{FLAG}=0$ for the PN, as recommended in the analysis threads from the {\it XMM-Newton} Science Operations Centre.\footnote{See \url{http://xmm.vilspa.esa.es/sas/7.1.0/documentation/threads/}.} Response files were created using the \verb#rmfgen# and \verb#arfgen# tasks, incorporating a correction for the flux missed by the finite extraction regions. The spectra were fitted with the eight spectral models described in \S~\ref{sec:light_curves}. The spectra for each of the three detectors were fitted simultaneously, allowing only an overall normalization factor to vary between detectors to account for differences in calibration. Using the best-fitting model for each source, an energy flux in the $0.3-7$ keV band was then computed. The individual fluxes for the three instruments were combined into a single flux result by taking the mean weighted with the inverse variance (i.e., squared error) of each individual flux.

\section{Observations and Source Detections}\label{sec:observations}

\subsection{NGC 6946}\label{sec:ngc6946}
NGC 6946 is one of the closest large spiral galaxies and is seen nearly face-on. It is classified by \cite{rc3} as type SAB(rs)cd. It has been extensively studied in almost all wavelength bands and has been shown to have strong nuclear starburst activity \citep{elmegreen1998b}. High star formation activity in its disk has also been observed \citep{sauty1998}, and in agreement with this NGC 6946 has been host to nine recorded supernovae in the last 91 years (three in the past 7 years), currently the highest known number of historical supernovae in any galaxy. The global star formation rate in NGC 6946 for stars in the mass range $2-60\textrm{ }M_{\sun}$ is estimated by \cite{sauty1998} to be $\sim4\textrm{ }M_{\sun}\textrm{ yr}^{-1}$, and \cite{crosthwaite2007} estimate the dynamical mass of the galaxy to be $\sim2\times10^{11}\textrm{ }M_{\sun}$.

The distance to NGC 6949 is given by \citet{devaucouleurs1979} as 5.1 Mpc and is listed as 5.5 Mpc in \citet{tully1988}, assuming a Hubble constant value of $75\textrm{ km s}^{-1}\textrm{ Mpc}^{-1}$. More recently, \citet{eastman1996} derive a distance of $5.7\pm0.7\textrm{ Mpc}$ based on a Type II supernova, and \citet{sharina1997} find a distance of $6.4\pm0.4\textrm{ Mpc}$ based on photometric observations of the brightest stars in the galaxy. Finally, \citet{karachentsev2000} estimate the distance to the NGC 6946 galaxy group (consisting of NGC 6946 and seven other small galaxies) to be $5.9\pm0.4\textrm{ Mpc}$, based on observations of the brightest blue giants in the galaxies. We choose to adopt a distance of 5.9 Mpc, as do \citet{holt2003}, \citet{schlegel2003}, and \citet{larsen2002}. The Galactic absorption column in the direction of NGC 6946 is $2.0\times10^{21}\textrm{ cm}^{-2}$ \citep{dickey1990,stark1992,kalberla2005}.

\begin{deluxetable*}{lccccccccc}[b]
\tablewidth{0pt}
\tabletypesize{\scriptsize}
\tablecaption{{\it XMM-Newton} EPIC Observations of NGC 6946\label{obs6946xmm}}
\tablehead{\multicolumn{2}{c}{} & \multicolumn{2}{c}{MOS-1} & \multicolumn{1}{c}{} & \multicolumn{2}{c}{MOS-2} & \multicolumn{1}{c}{} & \multicolumn{2}{c}{PN}\\
\cline{3-4} \cline{6-7} \cline{9-10}
\colhead{Start Date} & \colhead{Obs. ID} & \colhead{Exp.\tablenotemark{a}} & \colhead{Mode\tablenotemark{b}/} & \colhead{} & \colhead{Exp.} & \colhead{Mode/} & \colhead{} & \colhead{Exp.} & \colhead{Mode/}\\
\colhead{} & \colhead{} & \colhead{(ks)} & \colhead{Filter\tablenotemark{c}} & \colhead{} & \colhead{(ks)} & \colhead{Filter} & \colhead{} &  \colhead{(ks)} & \colhead{Filter}}
\startdata
2004 Jun 09 & 0200670101 & 3.95 & FF-ME && 4.05 & FF-ME && 2.50 & FF-ME \\
2004 Jun 13 & 0200670301 & 11.47 & FF-ME && 11.68 & FF-ME && 8.93 & FF-ME \\
2004 Jun 25 & 0200670401 & 7.69 & FF-ME && 8.49 & FF-ME && 4.81 & FF-ME \\
2006 Jun 02 & 0401360201 & 4.40 & FF-ME && 4.31 & FF-ME && 3.81 & FF-ME \\
2006 Jun 18 & 0401360301 & 5.28 & FF-ME && 5.60 & FF-ME && 3.58 & FF-ME \\
\enddata
\tablenotetext{a}{Good live exposure time.}
\tablenotetext{b}{Instrument modes: $\textrm{FF}=\textrm{full frame}$.}
\tablenotetext{c}{Filters: $\textrm{ME}=\textrm{medium filter}$.}
\tablecomments{Six additional {\it XMM-Newton} EPIC observations of NGC 6946 have been made, but those were not used due to high background rates leaving very little good exposure time.}
\end{deluxetable*}

NGC 6946 was first observed in X-rays with the {\it Einstein} satellite; three observations with the {\it Einstein} IPC were made in 1979--1981. Analyzing these observations, \citet{fabbiano1987} resolve the X-ray emission from the galaxy into two peaks. NGC 6946 was observed with the {\it ROSAT} PSPC in 1992, twice with the {\it ASCA} satellite in 1993 and 1994, and three times with the {\it ROSAT} HRI in 1994--96. \citet{schlegel1994b} analyzes the {\it ROSAT} PSPC observation and resolves emission from the galaxy into nine point sources, as well as finding evidence for diffuse emission. Analyzing the first of the three {\it ROSAT} HRI observations and the second of the two {\it ASCA} observations \cite{schlegel2000} detect 14 point sources and diffuse emission from the galaxy. Most recently, NGC 6946 has been observed five times with {\it Chandra} between 2001 and 2004, and 11 times with {\it XMM-Newton} in 2003--2006. The most detailed studies so far of the X-ray emission components of NGC 6946 are presented in a pair of papers based on the first of the five {\it Chandra} observations. \citet{holt2003} study the point source population and find 72 sources; the diffuse emission is extensively studied in \citet{schlegel2003}.

The {\it Chandra} observations of NGC 6946 are summarized in Table~\ref{obs6946chandra}. These are all ACIS-S full-frame imaging observations taken in Timed Exposure mode. The frame time was in all cases set at the standard value of 3.2 s, and events were telemetered in the Faint format. The observations span a period of about 39 months and have a cumulative good live exposure time of 172.5 ks. In each observation the majority of sources falls on the back-illuminated S3 chip, and a smaller portion falls on the front-illuminated S2 and/or S4 chips. None of the observations covers the whole $D_{25}$ isophote.

The data were analyzed as described in \S~\ref{sec:analysis}. Source detection was performed on a $14.8\arcmin\times14.1\arcmin$ rectangle encompassing the system. A circular region of radius $6\arcsec$ at the center of the galaxy (circle centered at $\mathrm{R.A.}=308.71857\degr$, $\mathrm{decl.}=60.153726\degr$) was excluded from further analysis, since the sources there are too crowded and embedded in too much localized diffuse emission for reliable background-subtracted count rates to be extracted. This central region seems to contain at least four and possibly six point sources. Outside this region a total of 90 reliable source detections down to luminosities of a few times $10^{36}\textrm{ ergs s}^{-1}$ (depending on the location) were found to be coincident with the galaxy (see Fig.~\ref{fig:dss}). The source catalog, including selected source parameters, is presented in Table~\ref{source_prop6946}. We choose to give photon fluxes for each source in each observation, as opposed to counts or count rates, since the latter two quantities are not directly comparable between different sources and observations due to variations in exposure and effective area.

An adaptively smoothed image based on the co-added {\it Chandra} exposure is shown in Figure~\ref{fig:xray}. The \verb#csmooth# task in CIAO was used to create smoothed images for the $0.3-1$, $1-2$, and $2-7$ keV bands. The same smoothing scales were used for all three images. The scales were calculated by adaptively smoothing a full-band image with a significance between 3.5 $\sigma$ and 5 $\sigma$ above the local background. Each of the three images was also exposure corrected using exposure maps smoothed with the same scales as the images. The images were then combined to create a color-coded full-band image with red representing the soft band, green the medium band, and blue the hard band. Finally, to enhance the point sources, the combined image was smoothed with a Gaussian with a kernel radius of three pixels.

Of the 90 sources, 65 were previously detected by \cite{holt2003} and 25 were not. \cite{holt2003} also report two sources within the central region excluded from our analysis, as well as one that falls outside the extent of the galaxy as we define it. Moreover, three of their 72 detected sources fall below our significance threshold for inclusion, and, finally, one of their sources almost certainly suffers from a typographical error in its position (sources 35 and 36 in their Table 1 have the same position to within an ACIS pixel size).

We associate five of our 90 detected sources with foreground stars (see Table~\ref{source_prop6946}). Comparing our source locations to the USNO-A2.0 Catalogue \citep{monet1998}, available online through VizieR, we find that five sources are coincident with foreground stars in the catalog to within $0.8\arcsec$. We estimate the likelihood of a chance coincidence to be low. For a crude test we shifted the source positions northeast by $7\arcsec$. This resulted in no matches with sources in the catalog to within $2.5\arcsec$.

Positions of sources were compared with two existing lists of SNRs in NGC 6946: a list of 27 optically identified SNRs from \citet{matonick1997} and a list of 35 radio-selected SNR candidates from \citet{lacey2001}. Interestingly, the two lists only have two sources in common. We find one counterpart to a SNR in the former list and five counterparts to SNR candidates in the latter list, as well as a counterpart to one of the two common sources. Two of the nine historical supernovae in the galaxy took place within the timespan covered by the {\it Chandra} observations, and they are both detected. In addition to these, two of the older historical supernovae are detected.

We estimate how many of our detected sources can be expected to be cosmic background sources (CBSs; primarily active galactic nuclei and star-forming galaxies) by using the results of \cite{bauer2004} from the {\it Chandra} Deep Fields. They give formulae estimating the number of CBSs per square degree of the sky above a given limiting flux in two energy bands, $0.5-2$ and $2-8$ keV. Performing source detection on a merged $0.5-2$ keV image from all five observations, 75 of our 85 nonforeground sources were detected with a \verb#wavdetect# significance level above 5 $\sigma$. In the $2-8$ keV band, 53 of the 85 sources were detected, five of which were not detected in the soft band. Since the total exposure varies widely over the extent of the galaxy, so does the limiting count rate for detection. For our soft band detections, we estimate that a typical limiting count rate is $\sim2\times10^{-4}\textrm{ counts s}^{-1}$ in the $0.3-7$ keV band, and $\sim4\times10^{-4}\textrm{ counts s}^{-1}$ for our hard band detections. We convert these count rates to energy fluxes in the soft and hard bands, respectively, assuming a power law with a spectral index of 1.4 \citep[as done by][]{bauer2004}. We take into account absorption both in our own Galaxy and in NGC 6946. Using the results of \cite{tacconi1986}, we estimate that a typical absorption column through NGC 6946 is $\sim1\times10^{21}\textrm{ cm}^{-2}$. Adding Galactic absorption, this yields a total absorption column for CBSs of $\sim3\times10^{21}\textrm{ cm}^{-2}$. Using our derived soft and hard band flux limits in the formulae of \cite{bauer2004}, we get 22.4 sources in the soft band over the angular area of the $D_{25}$ isophote of NGC 6946 (corresponding to 30\% of our 75 soft band detections), and 23.6 sources in the hard band, or 45\% of our 53 detections. Overall, we estimate that $\sim28$ of our total 85 detections in the $0.3-7$ keV band can be expected to arise from CBSs.

The {\it ROSAT} HRI and {\it XMM-Newton} EPIC observations of NGC 6946 are summarized in Tables~\ref{obs6946rosat} and~\ref{obs6946xmm}. A total of 11 {\it XMM-Newton} EPIC observations have been made, all of which suffer from periods of high background. Six of them are completely unusable and were therefore not analyzed further, and they are not included in Table~\ref{obs6946xmm}.

\subsection{NGC 4485/4490}\label{sec:ngc4485}

NGC 4485 and NGC 4490 are a closely interacting pair of galaxies, variously classified as spiral or irregular systems (see Fig.~\ref{fig:dss}). In \cite{rc3} NGC 4490 is classified as type SB(s)d, and the smaller companion NGC 4485 as type IB(s)m. Both galaxies show signs of tidal disruption. \citet{tully1988} gives separate distances of 7.8 Mpc (NGC 4490) and 9.3 Mpc (NGC 4485) to the two galaxies, assuming a Hubble constant value of $75\textrm{ km s}^{-1}\textrm{ Mpc}^{-1}$. These distances are adopted by \citet{roberts2000} and \citet{liu2005}. Such disparate distances are, however, unphysical for such a closely interacting system. \citet{roberts2002}, \citet{elmegreen1998a}, and \citet{read1997} therefore assume a distance of 7.8 Mpc to both galaxies. A consistent distance of 8 Mpc to the pair is adopted by \citet*{clemens2002}, \citet{clemens1998, clemens1999, clemens2000}, \citet{thronson1989}, and \citet{viallefond1980}, based on the suspected membership of the system in the CVn II cloud \citep{rc2}, and this is the value we used. The Galactic absorption column in the direction of NGC 4485/4490 is $1.8\times10^{20}\textrm{ cm}^{-2}$ \citep{dickey1990,stark1992,kalberla2005}.

\begin{figure*}
\includegraphics*[height=8cm]{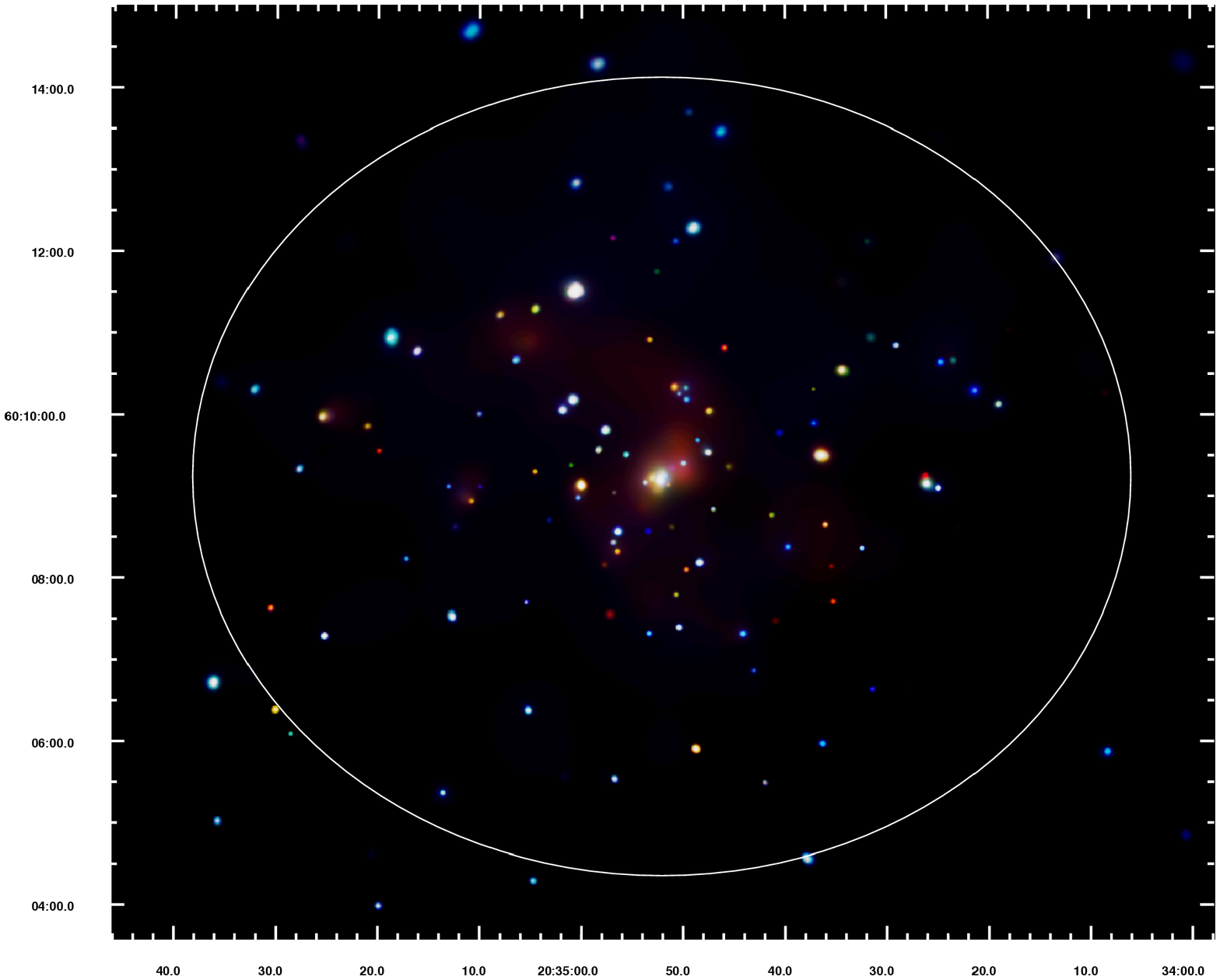}\label{fig:6946_xray}
\includegraphics*[height=8cm,trim= -30 0 0 0]{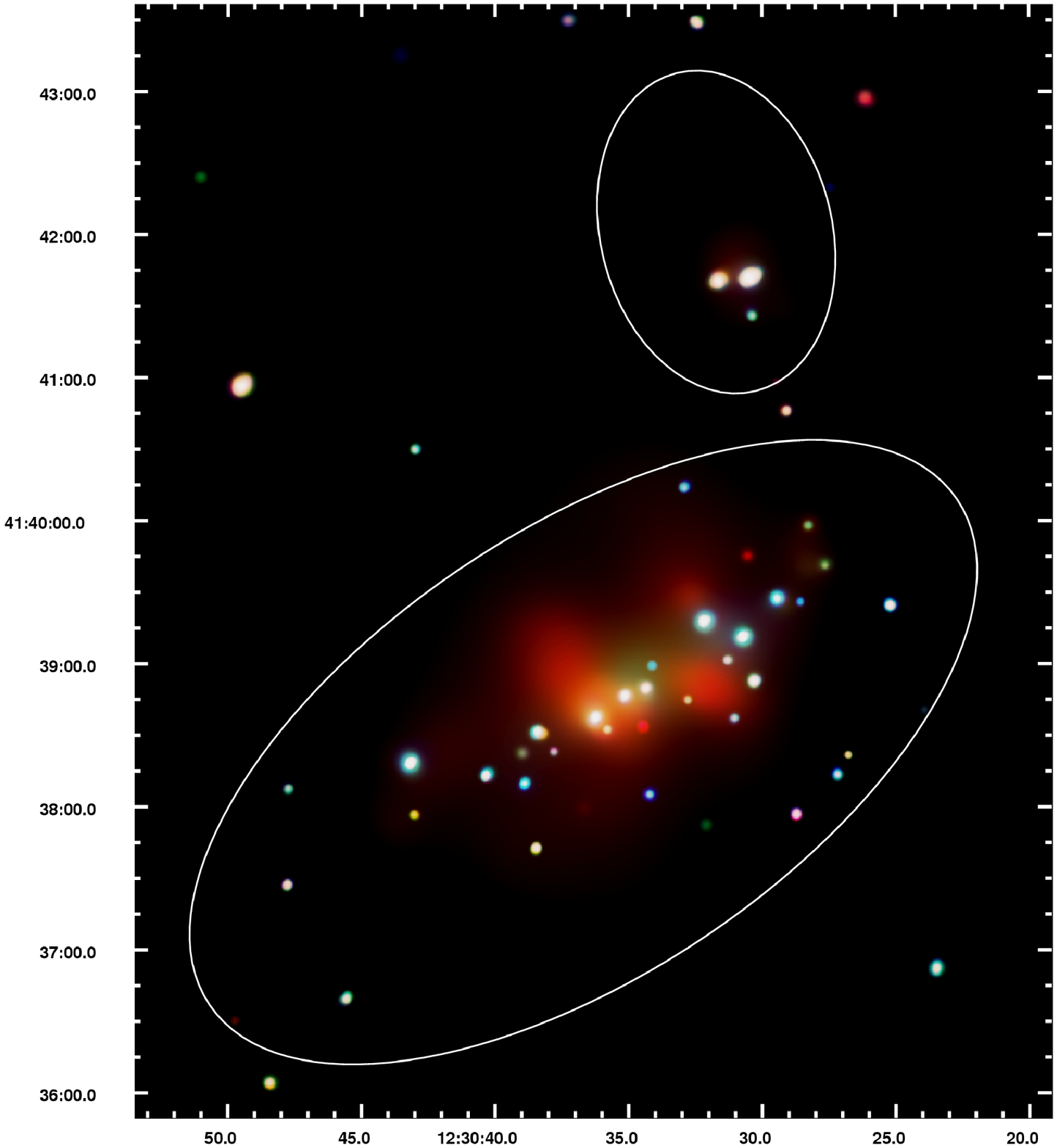}\label{fig:4485_xray}
\caption{Adaptively smoothed color-coded images based on the co-added {\it Chandra} exposures of NGC 6946 ({\it left}) and NGC 4485/4490 ({\it right}). Red corresponds to X-rays with energies $0.3-1$ keV, green to $1-2$ keV, and blue to $2-7$ keV.}\label{fig:xray}
\end{figure*}

\begin{deluxetable}{lcccc}
\tablewidth{0pt}
\tabletypesize{\scriptsize}
\tablecaption{{\it Chandra} Observations of NGC 4485/4490\label{obs4485chandra}}
\tablehead{\colhead{Start Date} & \colhead{Instrument} & \colhead{Obs. ID} & \colhead{Exp.\tablenotemark{a} (ks)} & \colhead{Data Mode\tablenotemark{b}}}
\startdata
2000 Nov 3  & ACIS-S & 1579 & 19.52 & F \\
2004 Jul 27 & ACIS-S & 4725 & 38.41 & VF \\
2004 Nov 20 & ACIS-S & 4726 & 39.59 & VF \\
\enddata
\tablenotetext{a}{Good live exposure time.}
\tablenotetext{b}{Telemetry formats: $\textrm{F}=\textrm{Faint}$, $\textrm{VF}=\textrm{Very Faint}$.}
\end{deluxetable}

NGC 4485/90 has been studied at radio, infrared, and optical wavelengths and shows evidence for enhanced star formation \citep{viallefond1980,klein1983,thronson1989}. \cite{clemens1999} estimate a star formation rate in NGC 4490 of $\sim4.7\textrm{ }M_{\sun}\textrm{ yr}^{-1}$ and a dynamical mass of $\sim1.6\times10^{10}\textrm{ }M_{\sun}$ within 8 kpc (the companion NGC 4485 being less massive by about a factor of 8). The system was first observed in X-rays with the {\it ROSAT} PSPC in 1991. Analyzing those data, \citet{read1997} detected four point-like sources, three in NGC 4490 and one in NGC 4485. The galaxy pair was subsequently observed three times with the {\it ROSAT} HRI in 1995 and 1996. Analyzing two of those observations, \citet{roberts2000} only found one additional point source, an apparently transient source coincident with NGC 4490. \citet{liu2005} analyzed all three {\it ROSAT} observations as part of their study of ULXs using the {\it ROSAT} HRI archive. They found the same sources as \citet{roberts2000}. Most recently, the system was observed three times with {\it Chandra} between 2000 and 2004 and once with {\it XMM-Newton} in 2002. Analyzing the first of the {\it Chandra} observations, \citet{roberts2002} found 31 point sources, 29 of those being coincident with NGC 4490, one coincident with NGC 4485, and one source located in a bridge between the galaxies.

The {\it Chandra} observations of the NGC 4485/90 system are summarized in Table~\ref{obs4485chandra}. These are all ACIS-S full-frame imaging observations taken in Timed Exposure mode. The frame time was in all cases set at the standard value of 3.2 s, and events were telemetered in either the Faint or Very Faint formats. The observations span a period of about 4 yr and have a cumulative good live exposure time of 97.5 ks. In all three observations all the sources coincident with NGC 4485/90 fall within the S3 chip.

The data were analyzed as described in \S~\ref{sec:analysis}. Source detection was performed on a $8.8\arcmin\times8.7\arcmin$ rectangle encompassing the system. A total of 38 reliable source detections down to a luminosity of $\sim1\times10^{37}\textrm{ ergs s}^{-1}$ were found to be coincident with the two galaxies. Of these, 33 are coincident with NGC 4490, three coincident with NGC 4485, and two more are located in the bridge between the galaxies (see Fig.~\ref{fig:dss}). The source catalog is presented in Table~\ref{source_prop4485}. An adaptively smoothed color-coded image based on the co-added {\it Chandra} exposure is shown in Figure~\ref{fig:xray}.

Of the 38 sources, 27 had previously been detected by \cite{roberts2002} and 11 had not been seen before. In addition, four of the sources reported in \cite{roberts2002} fell below our threshold of detection significance and are therefore not included in our tables. Comparing our source locations to the USNO-A2.0 Catalogue \citep{monet1998}, we find no close matches between the locations of foreground stars and our detected sources. There have been two historical supernovae in NGC 4490: SN 1982F, which is not detected, and recently SN 2008ax.

\begin{deluxetable}{lcc}[b!]
\tablewidth{0pt}
\tabletypesize{\scriptsize}
\tablecaption{{\it ROSAT} HRI Observations of NGC 4485/4490\label{obs4485rosat}}
\tablehead{\colhead{Start Date} & \colhead{Seq. ID} & \colhead{Exp.\tablenotemark{a} (ks)}}
\startdata
1996 Jun 18 & rh600855n00 & 26.70\\
1996 Nov 26 & rh600855a01 & 24.58\\
\enddata
\tablenotetext{a}{Good live exposure time.}
\tablecomments{One additional {\it ROSAT} HRI observation was made. We did not make use of it due to its shortness.}
\end{deluxetable}

We estimate how many of our detections can be expected to arise from CBSs using the same procedure as the one described for NGC 6946 in \S~\ref{sec:ngc6946}. We use the results of \cite{clemens1998} to estimate that a typical absorption column through the NGC 4485/90 system is $\sim3.5\times10^{21}\textrm{ cm}^{-2}$, giving a total absorption column of $\sim3.7\times10^{21}\textrm{ cm}^{-2}$ for CBSs behind NGC 4485/90. Applying the formulae of \cite{bauer2004} then yields an estimate of $\sim5$ for the expected number of CBSs among our 38 detected sources. For comparison, we note that \cite{roberts2002}, using the results of \cite{giacconi2001}, attribute $<4.5$ of their detections to CBSs.

The {\it ROSAT} HRI and {\it XMM-Newton} EPIC observations of NGC 4485/90 are summarized in Tables~\ref{obs4485rosat} and~\ref{obs4485xmm}. The first of the {\it ROSAT} observations is too short to be of significant use to this study, and it is not included in Table~\ref{obs4485rosat}.

\begin{deluxetable*}{lccccccccc}
\tablewidth{0pt}
\tabletypesize{\scriptsize}
\tablecaption{{\it XMM-Newton} EPIC Observation of NGC 4485/4490\label{obs4485xmm}}
\tablehead{\multicolumn{2}{c}{} & \multicolumn{2}{c}{MOS-1} & \multicolumn{1}{c}{} & \multicolumn{2}{c}{MOS-2} & \multicolumn{1}{c}{} & \multicolumn{2}{c}{PN}\\
\cline{3-4} \cline{6-7} \cline{9-10}
\colhead{Start Date} & \colhead{Obs. ID} & \colhead{Exp.\tablenotemark{a}} & \colhead{Mode\tablenotemark{b}/} & \colhead{} & \colhead{Exp.} & \colhead{Mode/} & \colhead{} & \colhead{Exp.} & \colhead{Mode/}\\
\colhead{} & \colhead{} & \colhead{(ks)} & \colhead{Filter\tablenotemark{c}} & \colhead{} & \colhead{(ks)} & \colhead{Filter} & \colhead{} &  \colhead{(ks)} & \colhead{Filter}}
\startdata
2002 May 27 & 0112280201 & 16.34 & FF-ME && 16.37 & FF-ME && 11.17 & EFF-ME \\
\enddata
\tablenotetext{a}{Good live exposure time.}
\tablenotetext{b}{Instrument modes: $\textrm{FF}=\textrm{full frame}$, $\textrm{EFF}=\textrm{extended full frame}$.}
\tablenotetext{c}{Filters: $\textrm{ME}=\textrm{medium filter}$.}
\end{deluxetable*}

\section{Properties of the Source Populations}\label{sec:properties}

\subsection{Flux Variability}

\begin{figure}[b]
\centerline{\includegraphics[width=5cm]{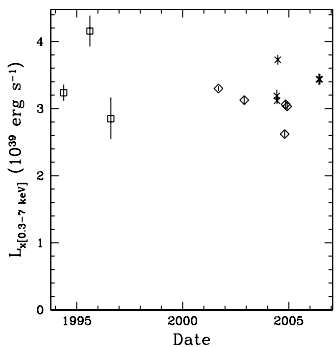}}
\caption{Long-term light curve of the ULX in NGC 6946 (CXOU~J203500.7+601130). Diamonds represent {\it Chandra} ACIS observations, crosses represent {\it XMM-Newton} EPIC observations, and squares represent {\it ROSAT} HRI observations.}\label{fig:lc_203500.74+601130.5}
\end{figure}

The variability properties of the sources in NGC 6946 are summarized in Table~\ref{var_prop6946}. Based on the source flux variability parameter $S_\mathrm{flux}$ defined in \S~\ref{sec:variability} we classify 25 of the 85 nonforeground sources, or 29\%, as variable on timescales of weeks, months, or years. Of these, 17 sources are seen to vary by more than a factor of 2, and we classify 11 sources (or 13\% of the 85) as transient candidates (two of those are recent supernovae). Evidence for short-term (hours) variability, as defined in \S~\ref{sec:variability}, is seen in four, or 5\%, of the 85 sources. Two of those also show long-term variability. Unsurprisingly, none of the sources associated with SNRs from the lists of \citet{matonick1997} and \citet{lacey2001} shows variability of any kind above a level of 2 $\sigma$, with the exception of the ULX (see \S~\ref{sec:ULX_6946}).

The variability properties of the sources in NGC 4485/90 are summarized in Table~\ref{var_prop4485}. We classify 15 of the 38 sources, or 39\%, as showing variability on timescales of months or years. Of these, 10 sources vary by more than a factor of 2, and four sources (or 11\% of the 38) are transient candidates. Evidence for short-term variability is seen in four of the 38 sources. All of those also show long-term variability.

Due to the limited number of counts within each observation for most of the sources showing short-term variability, it is very difficult (with one marginal exception) to draw any concrete conclusions about the nature of the variability (e.g., to distinguish between burst-like behavior and more continuous decline or rise in count rate). The brightest of these sources, CXOU J203500.1+600908 in NGC 6946, seems in the first {\it Chandra} observation to show a burst-like increase in count rate on a timescale of $\sim5-10$ ks superposed on a steady decline in count rate throughout the whole observation.

\begin{figure*}
\centering
\includegraphics*[width=15cm]{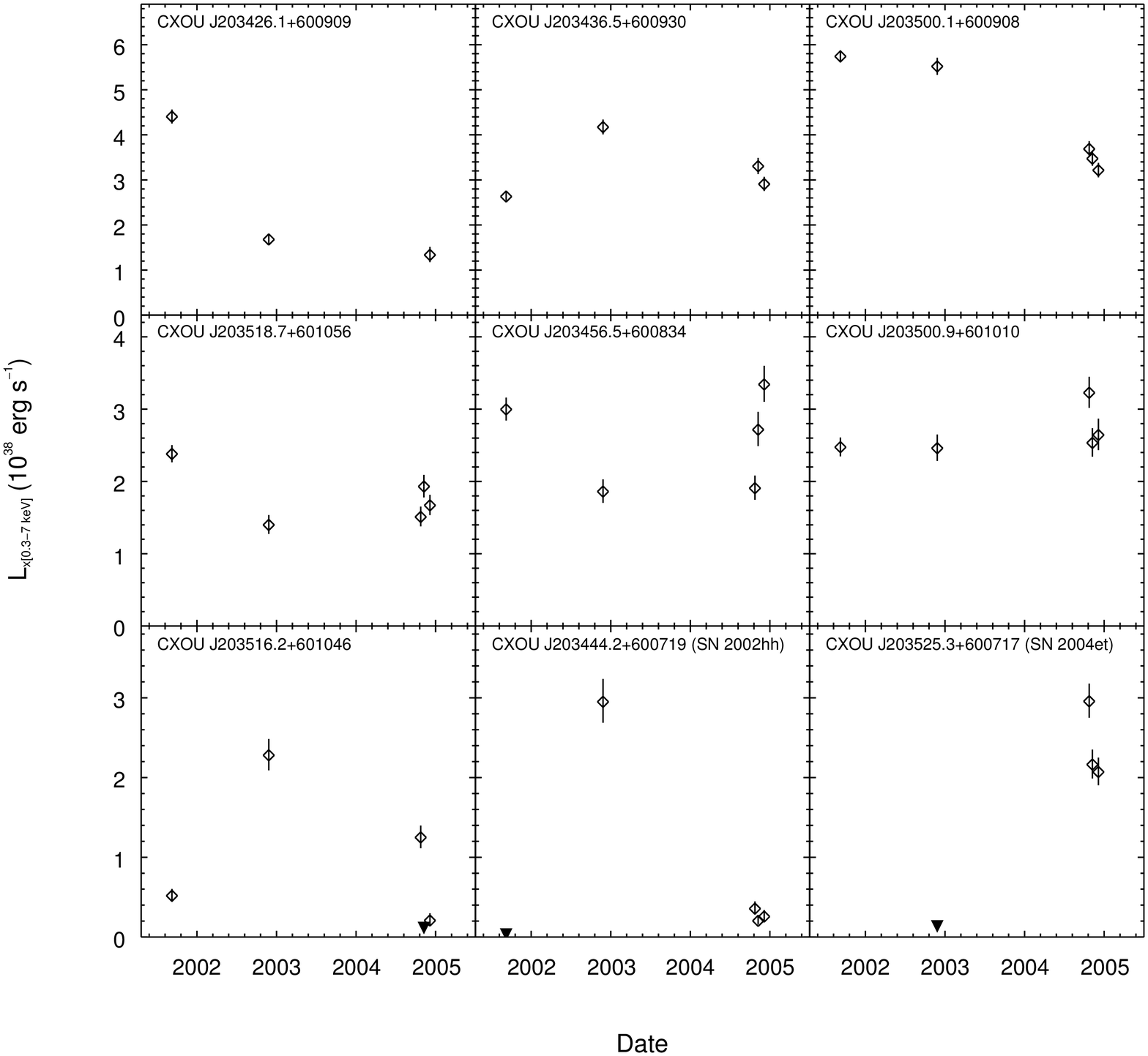}
\caption{Long-term {\it Chandra} light curves of nine of the most luminous sources in NGC 6946. Filled downward triangles represent upper limits with 3 $\sigma$ Gaussian confidence.}\label{fig:lc_6946}
\end{figure*}

Where possible, the baseline of the long-term light curves of the ULXs was extended by adding data points from {\it ROSAT} HRI and {\it XMM-Newton} observations as described in \S~\ref{sec:rosat_xmm}. Long-term light curves of the 10 brightest sources in NGC 6946 are shown in Figures~\ref{fig:lc_203500.74+601130.5} and~\ref{fig:lc_6946}, and of the 10 brightest sources in NGC 4485/90 in Figures~\ref{fig:lc_4485a} and ~\ref{fig:lc_4485b}. We emphasize that the error bars in the light curves contain only the 1 $\sigma$ Poisson errors in the counts. The {\it ROSAT} and {\it XMM-Newton} data are generally of much lower quality than the {\it Chandra} data, and comparisons of luminosities between different instruments should be regarded with some caution. Various factors (e.g., cross-calibration issues and uncertainties in instrument responses and in assumed spectral shapes for energy conversions) are likely to add to the errors. To give some idea of the magnitude of additional errors we note that the {\it Chandra} team estimates\footnote{See \url{http://cxc.harvard.edu/cal/}.} that present 1 $\sigma$ calibration uncertainties for ACIS are 0.3\% in the absolute energy scale, 5\% in the absolute effective area, and a few percent for the relative efficiency (i.e., for variations in quantum efficiency over the area of the detector). When choosing a spectral model to use for calculating ULX fluxes (as described in \S~\ref{sec:light_curves}) the scatter in flux values between several well-fitting models was typically of the order of a few percent. However, sometimes none of the eight simple spectral models considered provided a particularly good fit, in which case the errors added are likely greater. As for cross-calibration uncertainties between {\it Chandra} and {\it XMM-Newton}, we expect that the relative cross-calibration error between our {\it Chandra} and {\it XMM-Newton} fluxes should likely be $\lesssim10\%$ (H. Marshall, private communication, 2008).

The observed fractions here of $\sim30\%-40\%$ of sources showing flux variability is similar to what has been found in previous studies of late-type galaxies. \citet{kilgard2005} survey 11 nearby face-on spiral galaxies with two {\it Chandra} observations each and find that at least 27\% of the total number of sources exhibit variability on either long or short timescales. \citet{grimm2007} analyze three observations of M33 and find that 25\% of the detected sources exhibit long-term variability. They also classify 17\% of their sources as candidate transients, similar to what is seen in our galaxies. Overall, the abundant variability observed among the sources in NGC 6946 and NGC 4485/90 clearly points to source populations dominated by accreting XRBs.

It is interesting to compare the number of transient candidates seen in NGC 6946 and NGC 4485/90 to the number in our own Galaxy. Inspection of the data archives of the {\it Rossi X-Ray Timing Explorer's All-Sky Monitor} (RXTE ASM) reveals that on average $\sim4$ transients (including recurrent ones) with luminosities above $\sim10^{37}\textrm{ ergs s}^{-1}$ go off every year in the Galaxy (R. Remillard, private communication, 2008). We see 11 transient candidates in NGC 6946 and four in NGC 4485/90 (all with maximum luminosities above $10^{37}\textrm{ ergs s}^{-1}$). Due to the close monitoring of our Galaxy by the ASM, we would expect it to reveal a larger fraction of the actual number of transients among the X-ray sources than a handful of {\it Chandra} observations do in NGC 6946 and NGC 4485/90. However, a large fraction of the transient population of the Milky Way is presumably obscured by gas and dust. To put these numbers in context, we compare the estimated masses and star formation rates of these systems. A typical quoted value for the dynamical mass of the Milky Way within $\sim20$ kpc is a few times $10^{11}\textrm{ }M_{\sun}$ \citep[see, e.g.,][]{nikiforov2000}, and a recent estimate of its star formation rate is $\sim4\textrm{ }M_{\sun}\textrm{ yr}^{-1}$ \citep{diehl2006}. Corresponding numbers for NGC 6946 and NGC 4485/90 are given in \S\S~\ref{sec:ngc6946} and \ref{sec:ngc4485}. In short, the three systems are estimated to have similar star formation rates, and the Milky Way and NGC 6946 have comparable estimates for their mass, whereas NGC 4485/90 seems to be about an order of magnitude less massive.

\subsection{Individual Sources}

\subsubsection{Historical Supernovae in NGC 6946}

As mentioned in \S~\ref{sec:ngc6946}, NGC 6946 has been host to nine historical supernovae, the most recent one being SN 2008S. The last three of the five {\it Chandra} observations were aimed at Type~IIP SN~2004et and observed it 30, 45, and 72 days after the explosion (see light curve in Fig.~\ref{fig:lc_6946}). A multiwavelength study of this object in X-ray, optical, and radio is presented in \citet{misra2007}. Another {\it Chandra} study of SN~2004et is presented in \citet{rho2007}.

The second {\it Chandra} observation of NGC 6946, on 2002 November 25, was aimed at Type IIP SN~2002hh and observed it $\sim28$ days after the explosion, whose time is constrained to have been between 2002 October 26.1 and 31.1 UT \citep{li2002}. The measured luminosity at our adopted distance was $\sim3\times10^{38}\textrm{ ergs s}^{-1}$, with hardness ratios of $\textrm{HR1}=0.09\pm0.03$ and $\textrm{HR2}=0.78\pm0.12$. No source was detected at the position of the supernova in the first {\it Chandra} observation in 2001 September. In the last three {\it Chandra} observations, made around two years after the explosion, the source is seen to be emitting at a luminosity of $\sim3\times10^{37}\textrm{ ergs s}^{-1}$ (see light curve in Fig.~\ref{fig:lc_6946}). There is not a measurable difference in the soft color compared to the 2002 observation. However, the hard color exhibits significant softening between the 2002 and 2004 observations, being measured at $\textrm{HR2}=0.39\pm0.23$, $0.13\pm0.22$, and $-0.11\pm0.22$ in 2004. We note that SN 2004et exhibits similar spectral behavior, the soft color remaining constant but the hard color decreasing (see entries in Tables~\ref{spec_prop6946a} and~\ref{spec_prop6946b}), although the hard color is never seen to be as high as in the first observation of SN~2002hh. This is in agreement with the finding of \citet{misra2007} for SN 2004et that, while the $2-8$ keV flux decays, the $0.5-2$ keV flux remains roughly constant.

Type IIL SN 1980K fell outside the detector in the first {\it Chandra} observation but is seen to be emitting steadily at $\sim3\times10^{37}\textrm{ ergs s}^{-1}$ in the other four observations. We note that it was observed by the {\it Einstein} IPC 35 and 82 days after maximum light (which was $\sim15$ days after the explosion). From those observations \citet{canizares1982} report absorption corrected luminosities of $\sim7\times10^{38}\textrm{ ergs s}^{-1}$ and $\sim4\times10^{38}\textrm{ ergs s}^{-1}$ in the $0.2-4$ keV band (after scaling their values to our adopted distance). SN 1980K was also observed by the {\it ROSAT} PSPC in 1992 June. \cite{schlegel1994c} gives a luminosity estimate of $\sim2\times10^{37}\textrm{ ergs s}^{-1}$ in the $0.5-2$ keV range (scaled to our adopted distance). Type II SN 1968D is seen in all five {\it Chandra} observations and has a luminosity of $\sim2\times10^{37}\textrm{ ergs s}^{-1}$. The other four historical supernovae (SN~1917A, SN~1939C, SN~1948B, and SN~1969P) are not detected.

\begin{figure*}
\centerline{\includegraphics*[width=10cm]{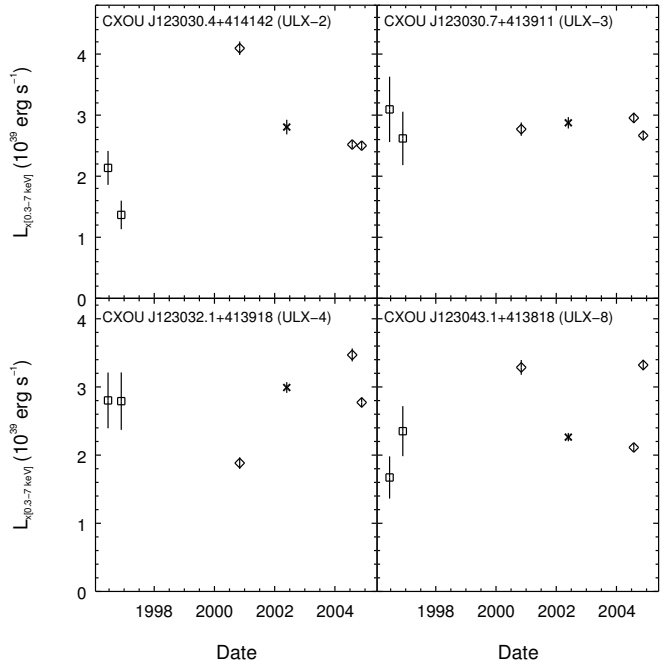}}
\caption{Long-term light curves of the four most luminous sources in NGC 4485/90. Diamonds represent {\it Chandra} ACIS observations, crosses represent {\it XMM-Newton} EPIC observations, and squares represent {\it ROSAT} HRI observations.}\label{fig:lc_4485a}
\end{figure*}

\subsubsection{ULX in NGC 6946}\label{sec:ULX_6946}

The only ultraluminous X-ray source in NGC 6946, often called MF16 \citep{matonick1997}, has been extensively studied. It was first detected with the {\it Einstein} IPC in 1979--1981 \citep{fabbiano1987} and was later observed with the {\it ROSAT} PSPC in 1992. Analyzing the {\it ROSAT} observation, \citet{schlegel1994a} derived an X-ray luminosity of $3.7\times10^{39}\textrm{ ergs s}^{-1}$ in the $0.5-2$ keV band (scaled to our adopted distance), and identified the source as a supernova remnant based on an optical counterpart discovered by \citet{blair1994}. The remnant appears to be quite evolved, perhaps $\sim10^3-10^4$ yr old \citep{blair1994,dunne2000,blair2001}. {\it Hubble Space Telescope} images presented by \citet{blair2001} reveal a three loop structure with an extent of $1.4\arcsec\times0.8\arcsec$ (corresponding to $39\times23$ pc at our adopted distance). Various models have been put forward to try to explain this unusually high luminosity from such an evolved SNR. \citet{blair2001} propose that ejecta from a young supernova might be interacting with the dense shell of an older SNR; \citet{dunne2000} suggest instead that a single supernova blast wave is interacting with a dense circumstellar nebula from a high-mass progenitor and that a hard spectral component in the {\it ROSAT} PSPC data comes from a pulsar wind nebula. \citet{schlegel2003}, analyzing the first {\it Chandra} observation, argue from spectral considerations that neither of these scenarios can explain the data. \citet{roberts2003} analyze the first two {\it Chandra} observations of the object and speculate that the luminosity of MF16 arises from a black hole X-ray binary (BHXRB) within the SNR. They argue this based on (1) the point-like nature of the X-ray source, (2) the close resemblance of the source spectrum to spectra from other ULXs thought to be BHXRBs, (3) the short-timescale (minutes) variability of the hard component ($1.1-10$ keV) of the flux, and (4) the long-timescale drop of $\sim13\%$ in count rate between the first and second {\it Chandra} observations ($\sim3\%$ thereof stemming from a degradation in effective area).
Our K-S test does not detect short-timescale variations in the full $0.3-7$ keV band for any of the {\it Chandra} observations. We do, however, see definite long-term variability in the {\it Chandra} data (see light curve in Fig.~\ref{fig:lc_203500.74+601130.5}). There is a drop in luminosity of $\sim5\%$ between the first and second observations, and then a much larger drop between the second and third observations; the luminosity in the third observation is $\sim79\%$ of the original first observation value. Two weeks later, in the fourth {\it Chandra} observation, the luminosity has again increased to $\sim93\%$ of its original value. We also see clear evidence of variability in the {\it XMM-Newton} data, most notably a $\sim20\%$ jump in luminosity in the 12 days between two observations in 2004 June. Finally, we see variations in the soft color with a significance of 2.9 $\sigma$ (see Fig.~\ref{fig:color-lum}). These spectral variations do not seem to have a clear correlation with flux. There is spectral softening associated with the drop in flux between the second and third {\it Chandra} observations, but there is no detectable hardening when the flux subsequently increases again. Overall, this observed variability considerably strengthens the case for an accreting object as the source of at least part of the X-ray emission from MF16.

\begin{figure*}
\centerline{\includegraphics*[width=14cm]{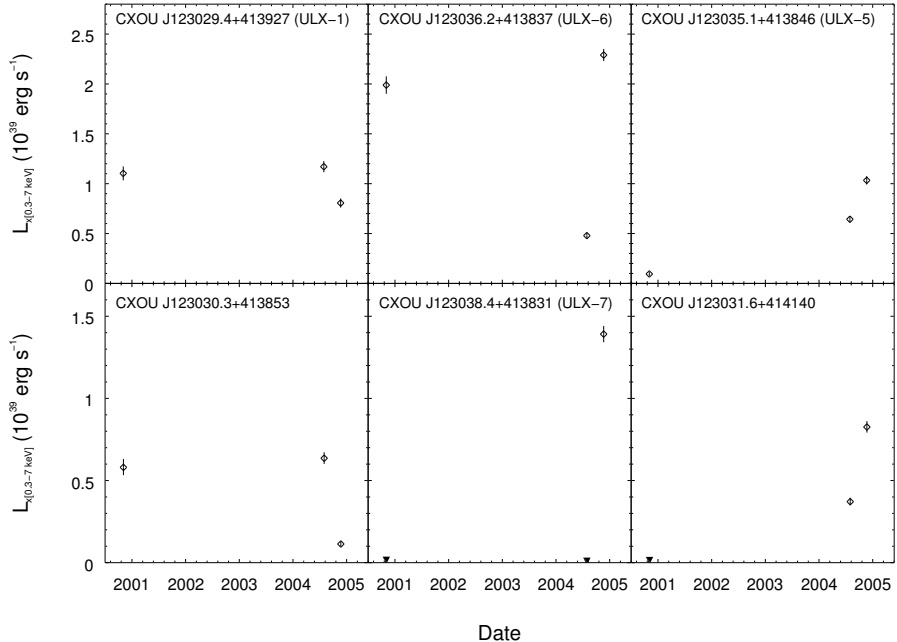}}
\caption{Long-term {\it Chandra} light curves of six of the most luminous sources in NGC 4485/90. Filled downward triangles represent upper limits with 3 $\sigma$ Gaussian confidence.}\label{fig:lc_4485b}
\end{figure*}

\subsubsection{ULXs in NGC 4485/4490}

The NGC 4485/90 system contains a high number of ULXs, a total of eight. In a comprehensive survey of nearby galaxies observed with the {\it ROSAT} HRI, \citet[][see also \citealt{liu2005}]{liu2006} find an average rate of $\sim0.5$ ULXs per galaxy. The rate is higher when only considering late-type galaxies and still higher for starburst galaxies ($\sim1.0$).
Considerably higher numbers, however, are not unusual for interacting galaxies \citep[see, e.g.,][]{brassington2007}, an extreme case being the Antennae, with 14 known ULXs \citep{zezas2006}.

Seven of the ULXs in the NGC 4485/90 system are in NGC 4490 and one is in NGC 4485. We numerate them here in order of right ascension (see Table~\ref{source_prop4485}). Two of them, ULX-5 and ULX-7, have not been identified as being ultraluminous before, their luminosities having entered the ULX range only in the most recent {\it Chandra} observation.

The maximum luminosities of the ULXs range from $1.0\times10^{39}\textrm{ ergs s}^{-1}$ to $4.1\times10^{39}\textrm{ ergs s}^{-1}$. These sources show a lot of variability. Long-term flux variability above the 3 $\sigma$ level is exhibited by all but one of them (see light curves in Figs.~\ref{fig:lc_4485a} and~\ref{fig:lc_4485b}). ULX-7 is a transient, ULX-5 varies by a factor of $\sim11$ and ULX-6 by a factor of $\sim5$. The others vary by less than a factor of 2. ULX-5 shows short-term flux variability. Two sources, ULX-6 and ULX-8, have significant variations in their hard color. In both cases, there is a clear transition from a harder higher luminosity state to a softer lower luminosity one and back (see Fig.~\ref{fig:color-lum}).

\citet{vazquez2007} use the Infrared Spectrograph on the {\it Spitzer Space Telescope} to do mid-infrared spectral diagnostics on six of the ULXs (all except ULX-5 and ULX-7). They find that five of those show ionization features that they claim indicate accretion, whereas the sixth source, ULX-1, seems to be more consistent with a SNR. As pointed out by \citet{roberts2002}, ULX-1 is coincident with a radio source, FIRST J123029.4+413927, which also points towards a SNR origin for ULX-1. However, we see a $\sim30\%$ drop in luminosity in the four months between the second and third {\it Chandra} observations---behavior that is more typical of an accreting XRB than a SNR. This is reminiscent of the ULX in NGC 6946, which has also been associated with a SNR but whose X-ray variability indicates an accreting XRB.

The abundance of variability among the ULXs in NGC 4485/90 is similar to what is seen in the Antennae. \citet{zezas2006} find that two of the ULXs in the Antennae show variability on a timescale of a few days. Twelve out of 14 ULXs show long-term variability, one being a transient, and four of the ULXs occasionally have luminosities below the ULX limit. In NGC 4485/90 we find that four of the eight ULXs have a luminosity below $10^{39}\textrm{ ergs s}^{-1}$ in at least one of the three {\it Chandra} observations. It is therefore clear that a single observation of a galaxy is by no means guaranteed to discover the galaxy's full ULX population. Surveys such as that of \citet{liu2006}, where many of the galaxies are only observed once or twice, can only give a lower limit to the average number of sources in a galaxy with the potential of being ultraluminous at one time or other.

\subsection{Hardness Ratios}

\subsubsection{Variability}

\begin{figure*}
\centering
\includegraphics[width=5.5cm]{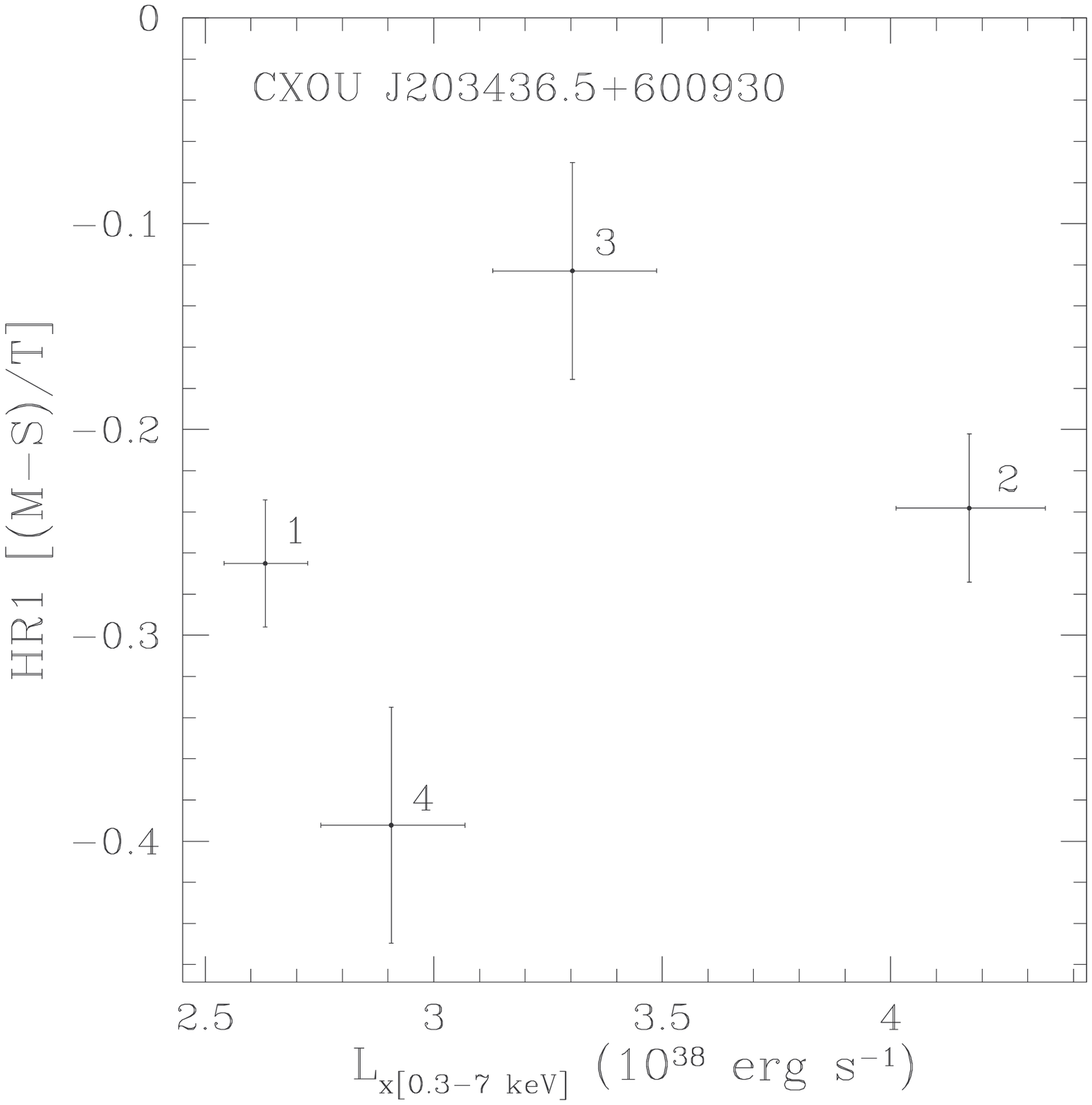}
\includegraphics[width=5.5cm]{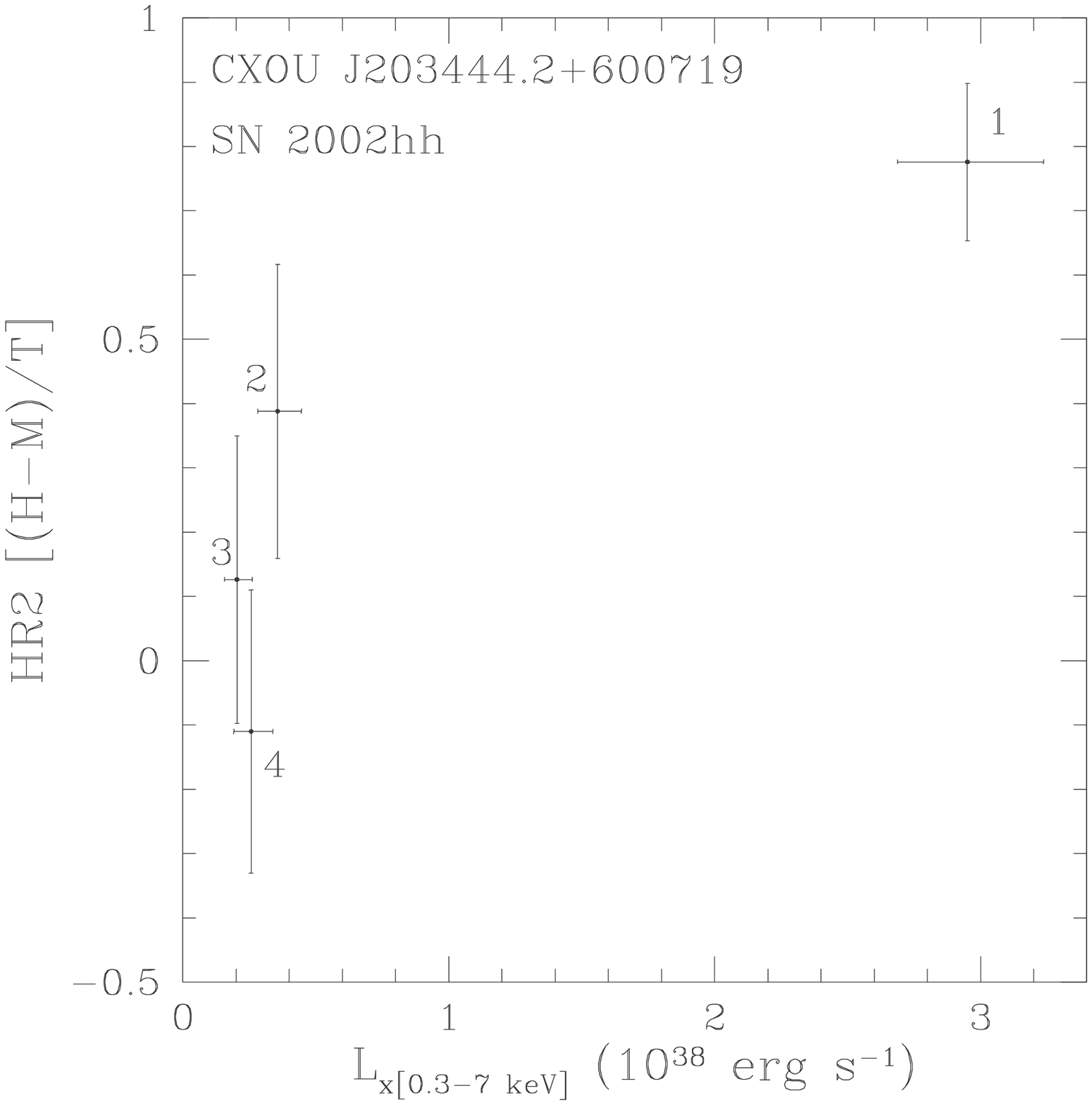}
\includegraphics[width=5.5cm]{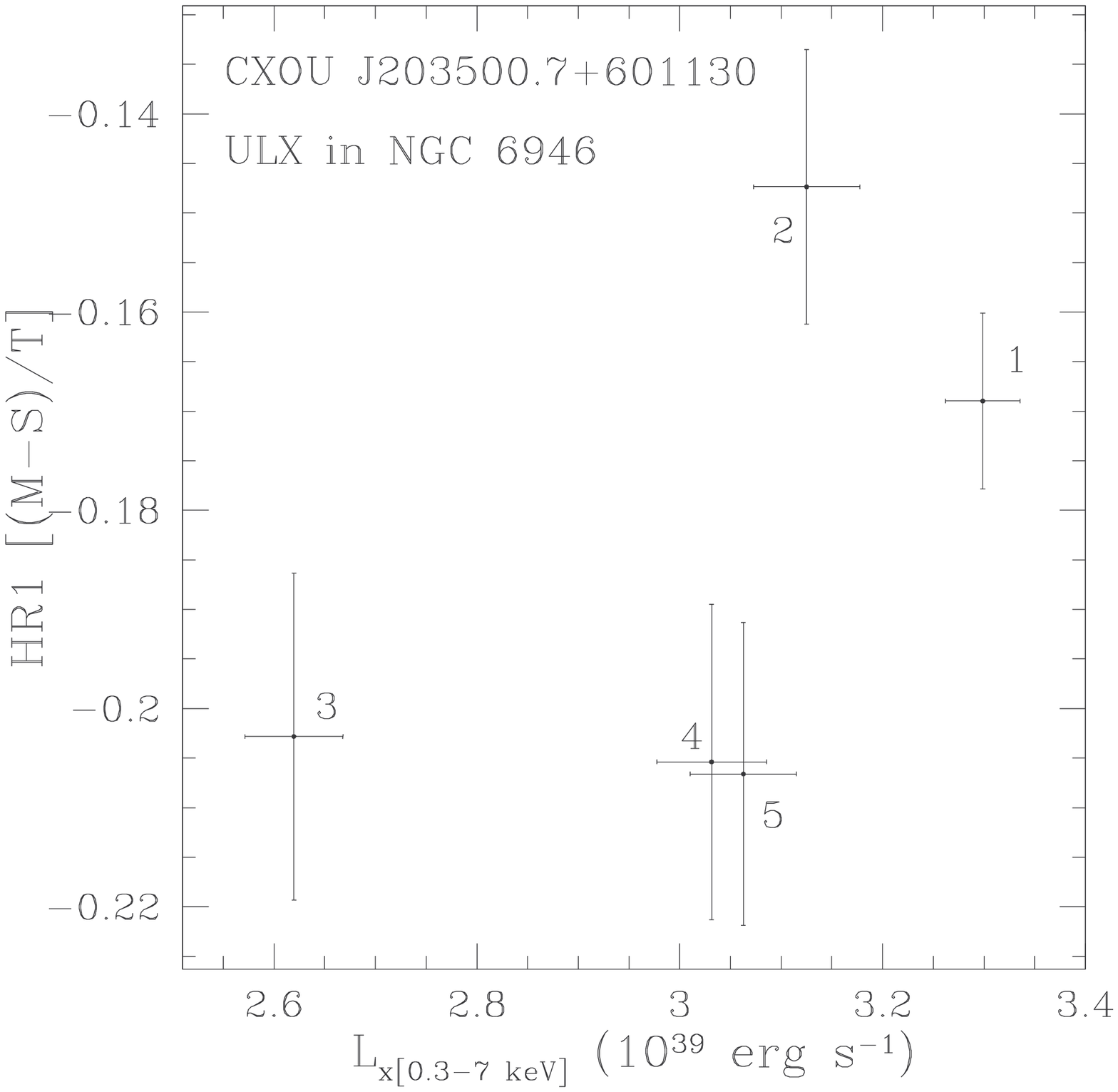}
\includegraphics[width=5.5cm]{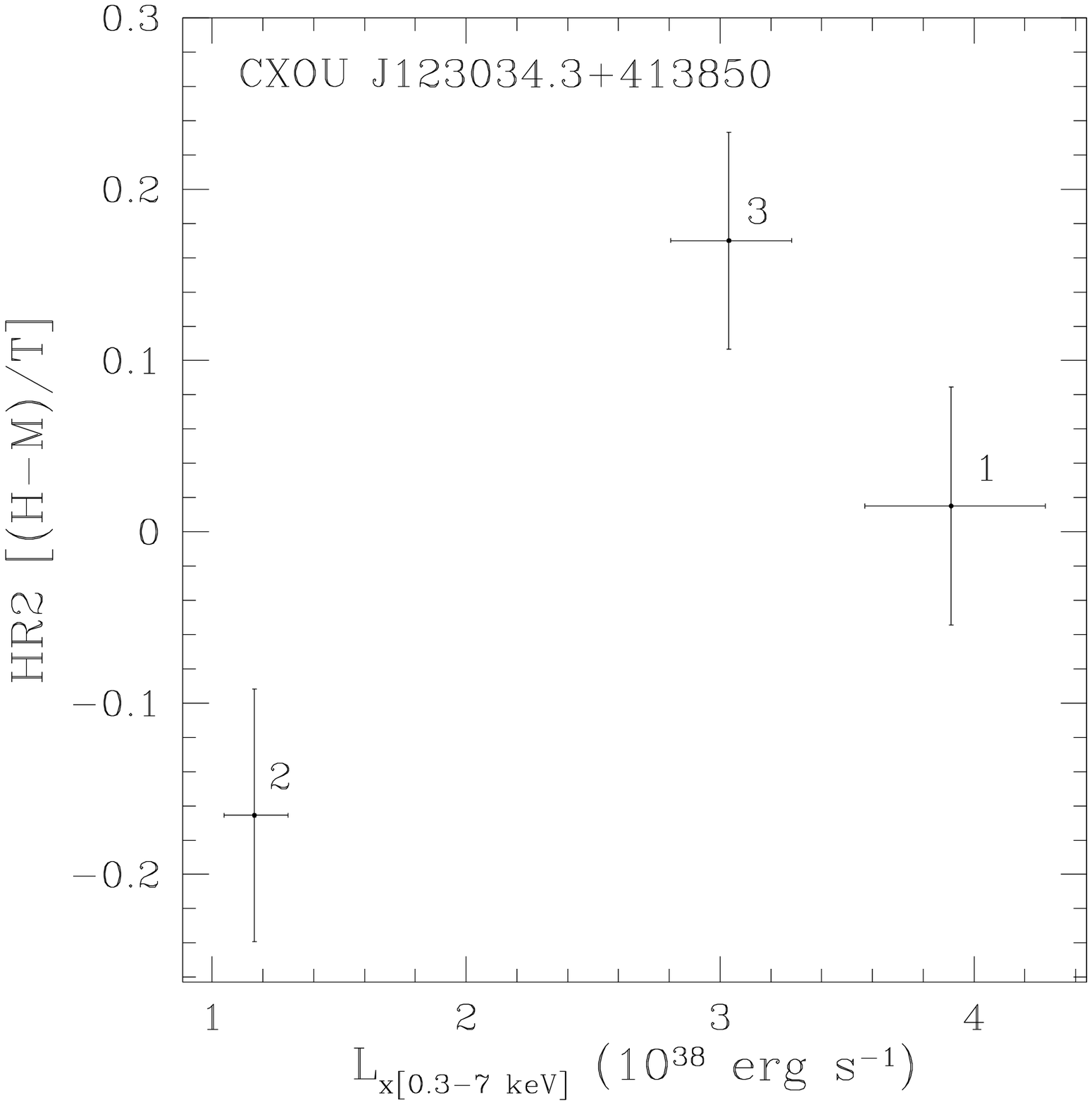}
\includegraphics[width=5.5cm]{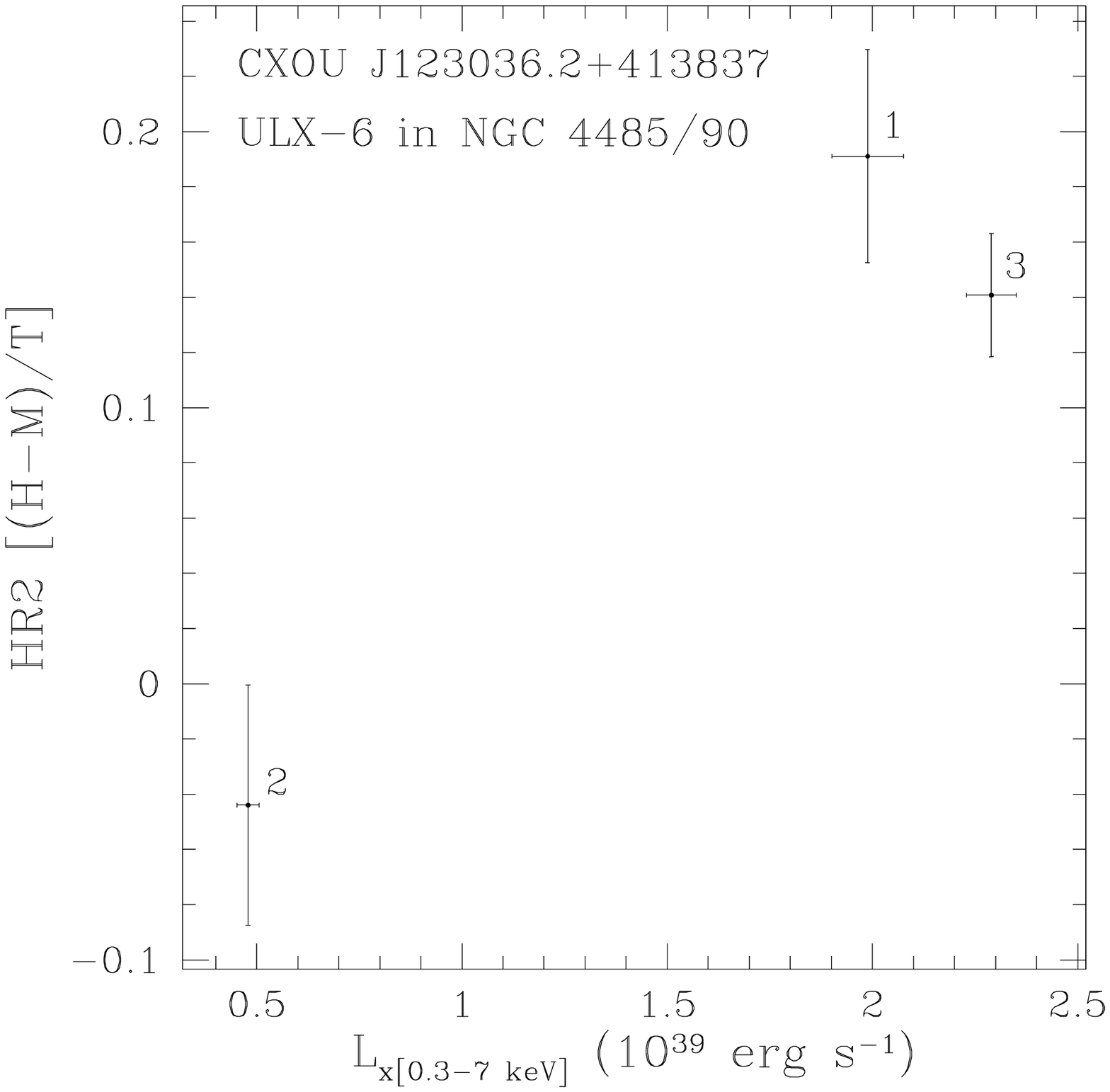}
\includegraphics[width=5.5cm]{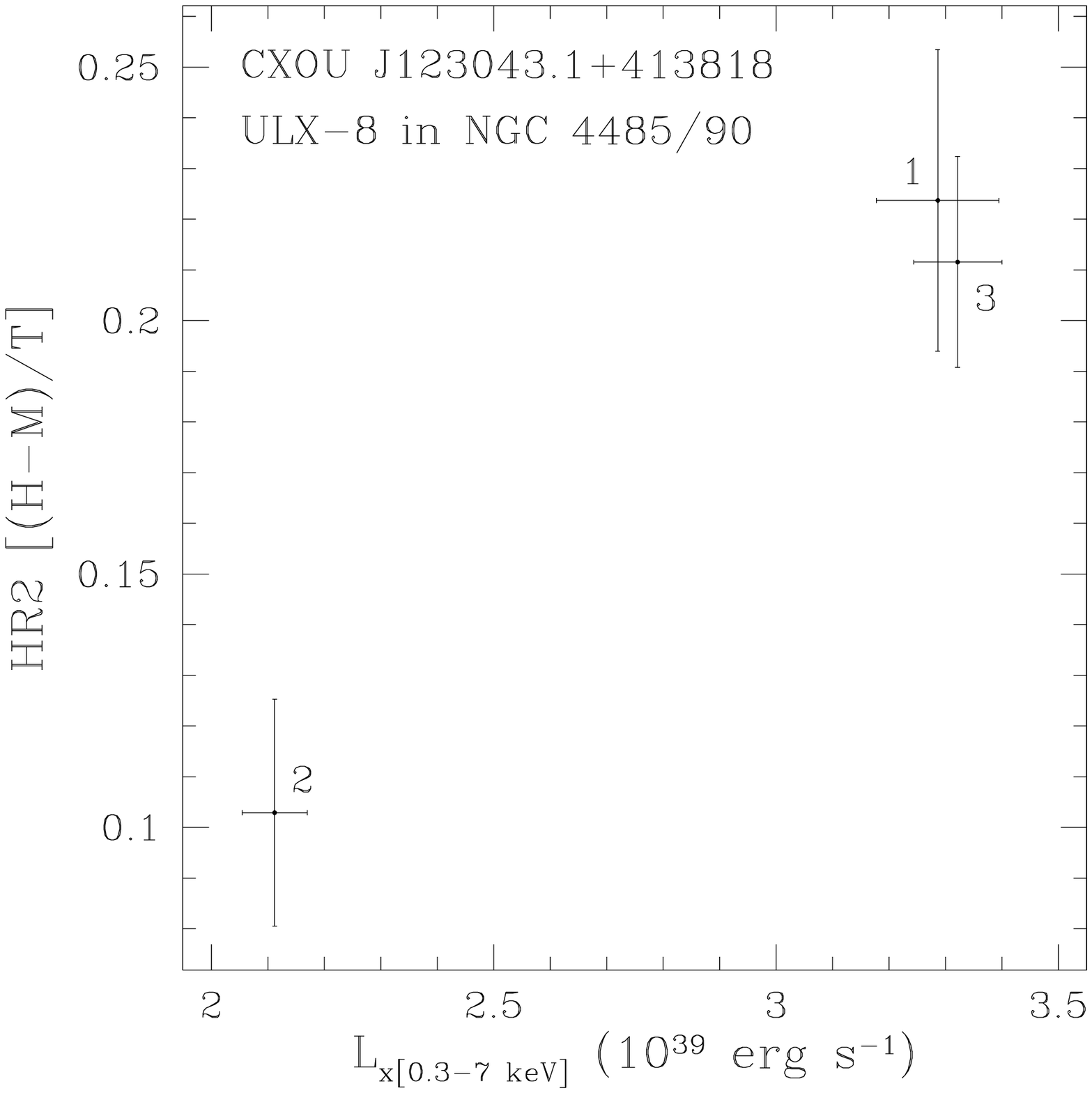}
\caption{Soft or hard color as a function of luminosity for sources that show significant variation in their hardness ratios. Numbers next to data points indicate their order in time.}\label{fig:color-lum}
\end{figure*}

The hardness ratios of the sources in NGC 6946, both for each observation and for the co-added exposure, are given in Tables~\ref{spec_prop6946a} and~\ref{spec_prop6946b}. The hardness ratios for the NGC 4485/90 sources are similarly given in Table~\ref{spec_prop4485}. Significance parameters for variability in the hardness ratios between observations are given in Tables~\ref{var_prop6946} (NGC 6946) and~\ref{var_prop4485} (NGC 4485/90). Variability above  3 $\sigma$ in either of the ratios is only detected in two of the sources in NGC 6946 and three of the sources in NGC 4485/90. In addition, the ULX in NGC 6946 shows a 2.9 $\sigma$ variation in its hard color. We note that a much larger fraction of the sources probably undergoes some spectral variation, but most of the sources have too few counts for the hardness ratios to be very sensitive to spectral changes. Color-luminosity plots for the sources exhibiting significant hardness ratio variability are shown in Figure~\ref{fig:color-lum}. Four of these sources (one of those being SN 2002hh) seem to show a clear correlation between higher luminosity and increased hardness, whereas the other two show more complicated behavior. Previous studies of external galaxies have routinely found sources with spectral variability, in some cases with spectral behavior reminiscent of Galactic XRBs; flux-color transitions have also been observed in ULXs \citetext{\citealp{fabbiano2006b} and references therein}. \citet{zezas2006} find signs of variability in hardness ratios of 21 of their $\sim70$ detected sources in the Antennae; some of these are ULXs. As in our case, they find that the variability does not follow the same pattern for all sources. They see sources that become harder with higher luminosity, as we do, but also sources that soften with increased luminosity. In addition, some of the sources show no regular pattern, or no significant variation in luminosity.

\subsubsection{Supersoft Sources}

Two sources have especially soft spectra and are strong candidates for supersoft sources \citep[SSSs; see, e.g.,][]{distefano2003}. CXOU J203426.2+600914 in NGC 6946 has all of its 68 source counts below 1 keV, and CXOU J123034.5+413834 in NGC 4490 has 32 of its total 35 source counts below 1 keV, 25 counts being below 0.65 keV. Both sources have a luminosity of $\sim9\times10^{36}\textrm{ ergs s}^{-1}$, and neither one shows evidence for variability.

\subsubsection{Spectral Classification}\label{sec:spec_class}

\begin{figure}
\centering
\includegraphics*[height=8.9cm,trim= 50 10 0 20]{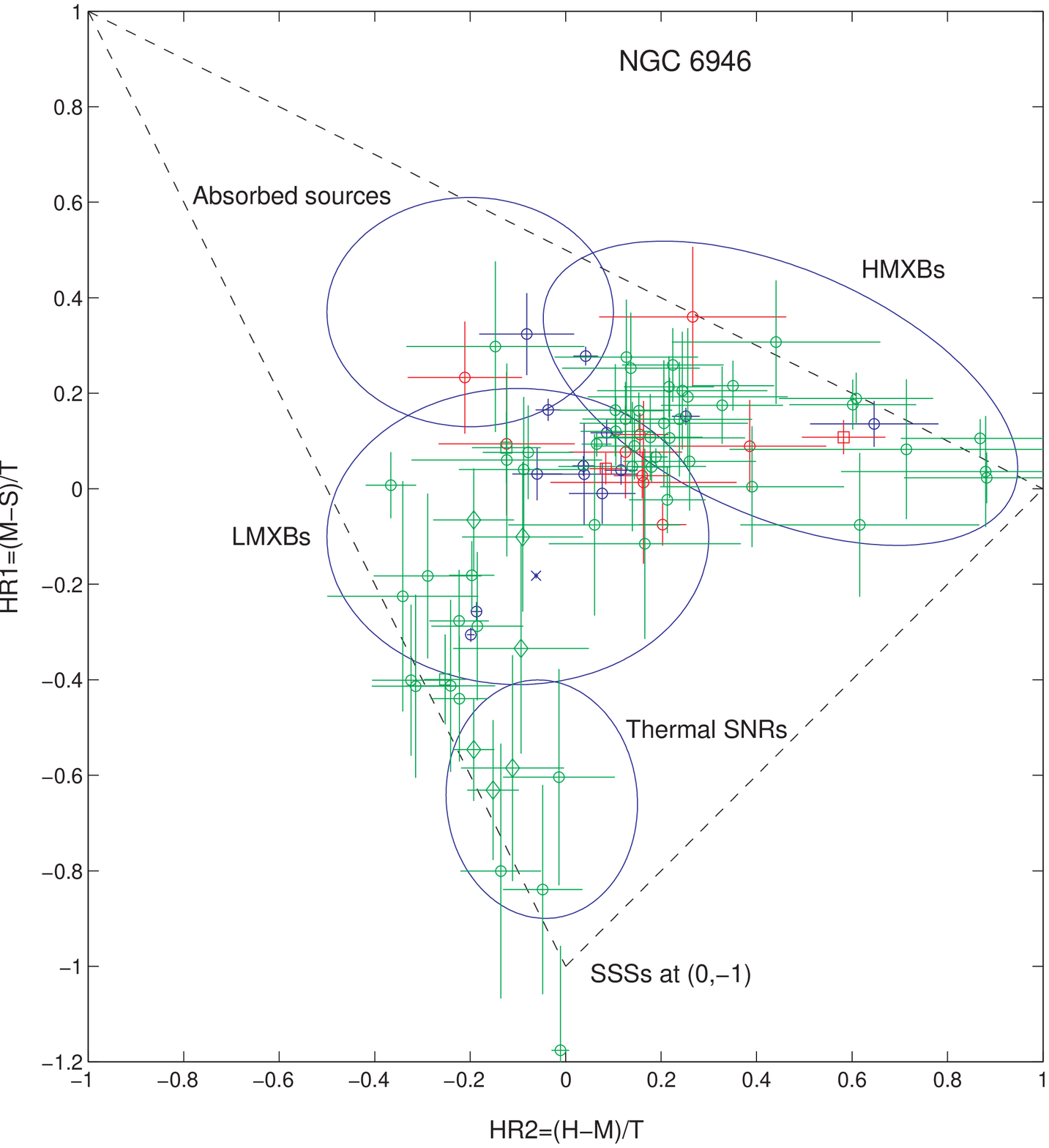}
\includegraphics*[height=8.9cm,trim= 50 10 0 40]{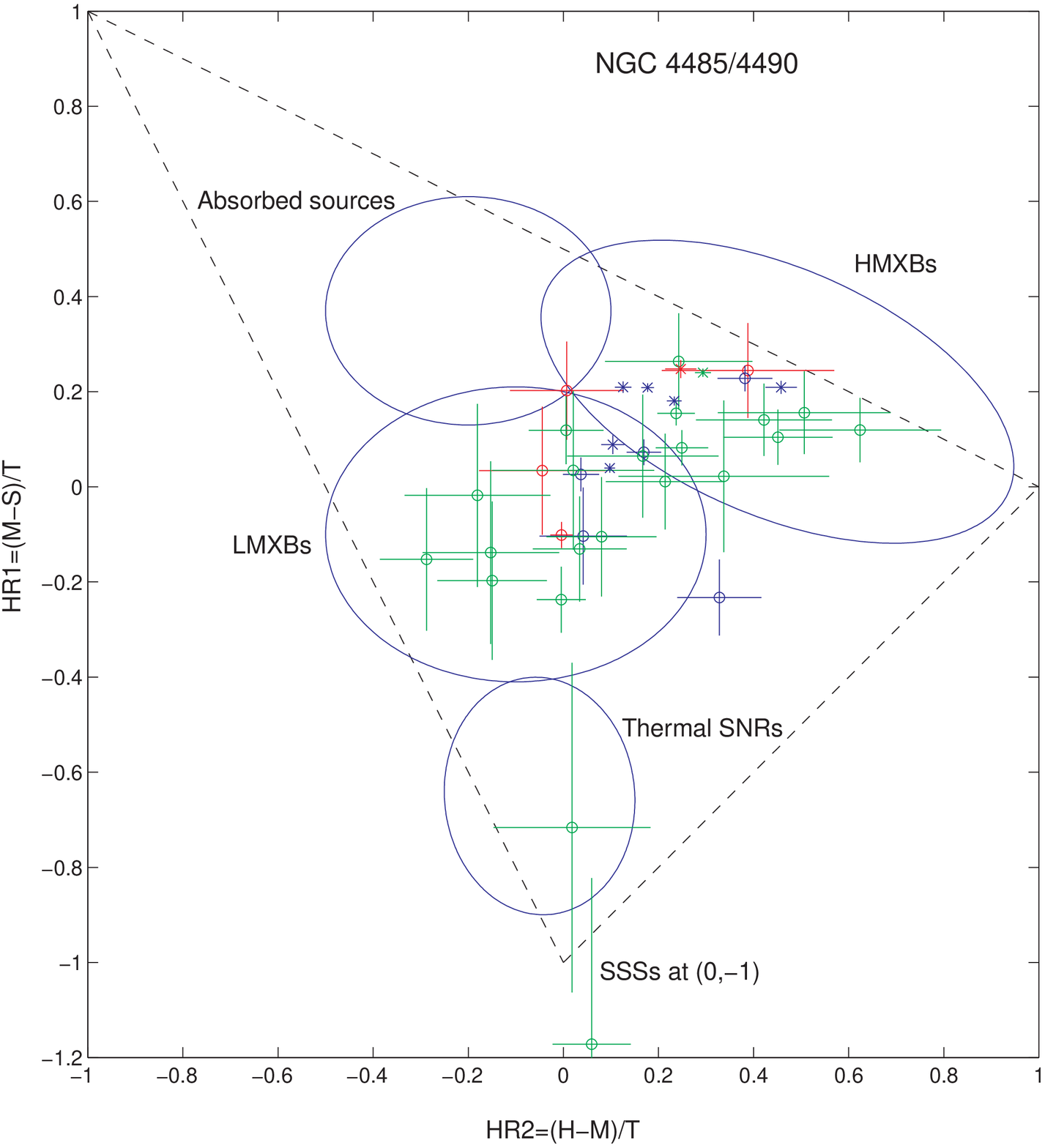}
\caption{Color-color diagrams for the sources in NGC 6946 and NGC 4485/4490 based on the co-added {\it Chandra} exposures. The classification scheme of \citet{prestwich2003} is overlaid. Historical supernovae are represented with boxes, SNRs with diamonds, ULXs with crosses, and other sources with circles. Blue indicates that long-term variability is detected, green that it is not, and red denotes transient candidates. The triangle enclosed by the dashed lines represents the physically allowed region where net counts in all energy bands are non-negative. Due to background subtraction, faint sources can fall outside this region.}\label{fig:spec}
\end{figure}

Color-color diagrams for the sources in NGC 6946 and NGC 4485/90, based on the co-added exposures, are shown in Figure~\ref{fig:spec}. The fact that most of the sources do not show significant variation in their hardness ratios between observations provides justification for using hardness ratios calculated from the co-added exposure to describe the source populations' spectral properties. The classification scheme of \citet{prestwich2003} discussed in \S~\ref{sec:hardness_ratios}, is overlaid. As mentioned before, although a source's position in a color-color diagram does not give conclusive information about the nature of the source, the overall distribution of sources should give us some information about the population as a whole. Also indicated in the color-color diagrams is the physically allowed region, covering all combinations of HR1 and HR2 possible if the net counts in all the energy bands (S, M, H, T; see \S~\ref{sec:hardness_ratios}) are non-negative. Due to background subtraction in cases of very few source counts, some of the hardness ratios fall outside this region. They are, however, all consistent with the physically allowed region within their 1 $\sigma$ errors.

\citet{prestwich2003} note that, while thermal SNRs are in general expected to have $-0.9<\textrm{HR1}<-0.4$, as indicated in the classification scheme, SNRs whose spectra are dominated by nonthermal components will have somewhat harder spectra and are likely to be located in the LMXB or even absorbed sources part of the color-color diagram. We see in Figure~\ref{fig:spec} that the locations of the sources associated with SNRs in NGC 6946 are in good agreement with this.

For NGC 4485/90, all but two of the 38 sources fall in the XRB regions of the color-color diagram, with roughly equal portions falling in the LMXB and HMXB regions. In NGC 6946 there seems to be a much larger SNR component to the population. In addition to the four detected historical supernovae and the six sources associated with optically or radio identified SNRs, $\sim5$ sources have colors consistent with thermal SNRs. The rest of the sources in NGC 6946 (apart from the SSS candidate) have colors consistent with XRBs, again with roughly equal portions in the LMXB and HMXB regions. We note that a large portion of the sources has colors consistent with both regions. The large presence of HMXBs in these systems, indicated by the color-color diagrams, is not surprising given the high level of star formation observed.

All the ULXs have colors consistent with XRBs. The ULX in NGC 6946 falls firmly in the LMXB region, whereas six of the ULXs in NGC 4485/90 have colors more consistent with HMXBs. Variable and transient sources are equally divided between the LMXB and HMXB regions in NGC 4485/90. For NGC 6946, these sources seem to slightly favor the LMXB region, but this is hardly a significant difference. The most variable sources (those that vary by a factor of 10 or more) are also seen to be roughly equally divided between the LMXB and HMXB regions in both galaxies.

\subsection{Luminosity Functions}\label{sec:XLF}

We construct X-ray luminosity functions (XLFs) for each observation, as well as the co-added exposure, of both NGC 6946 and NGC 4485/90. We use luminosities from the full $0.3-7\textrm{ keV}$ band. In the case of NGC 6946, where parts of the galaxy fell outside the {\it Chandra} detectors in each observation, only sources that fell within the detectors in all five observations were considered. Hence, 24 of the 85 nonforeground sources were discarded. This was done to make sure that when comparing the XLFs from different observations we are comparing the same source population.

To avoid any effects of incompleteness at the low-luminosity end of the XLF, we set a conservative completeness limit for both galaxies and consider only the portion of the XLF above that limit. To facilitate comparisons between different observations we use the same completeness limit for all observations of each galaxy. We fit a simple power law to the unbinned distribution for each observation using the Cash statistic \citep{cash1979}. In order to test the null hypothesis that the XLFs of different observations of the same galaxy are derived from the same parent distribution, we compare the XLFs in pairs using a two-sided Kolmogorov-Smirnov test.

\begin{deluxetable}{cc}
\tablewidth{0pt}
\tabletypesize{\scriptsize}
\tablecaption{Power-law Indices of the XLFs of NGC 6946\label{xlf6946}}
\tablehead{\colhead{Obs.} & \colhead{$\alpha$\tablenotemark{a}}}
\startdata
1043 & $0.61\pm0.13$\\
4404 & $0.63\pm0.10$\\
4631 & $0.67\pm0.11$\\
4632 & $0.72\pm0.13$\\
4633 & $0.67\pm0.12$\\
Co-added & $0.73\pm0.14$\\
\enddata
\tablenotetext{a}{Best-fit cumulative slope. Uncertainties given are estimated 90\% confidence intervals.}
\end{deluxetable}

\begin{figure}
\centering
\includegraphics*[height=8.5cm,trim= 0 0 0 40]{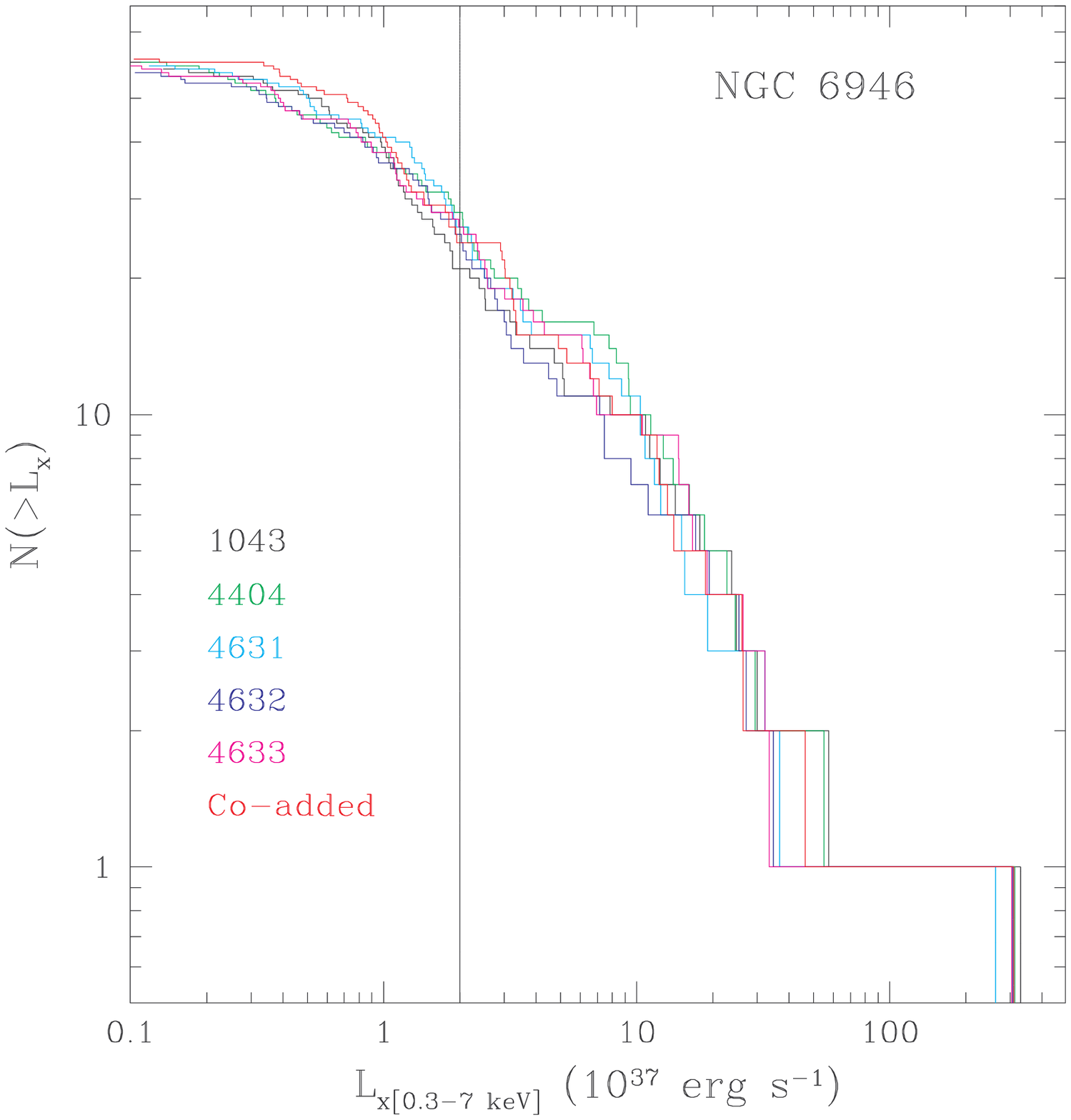}
\includegraphics*[height=8.5cm,trim= 0 0 0 30]{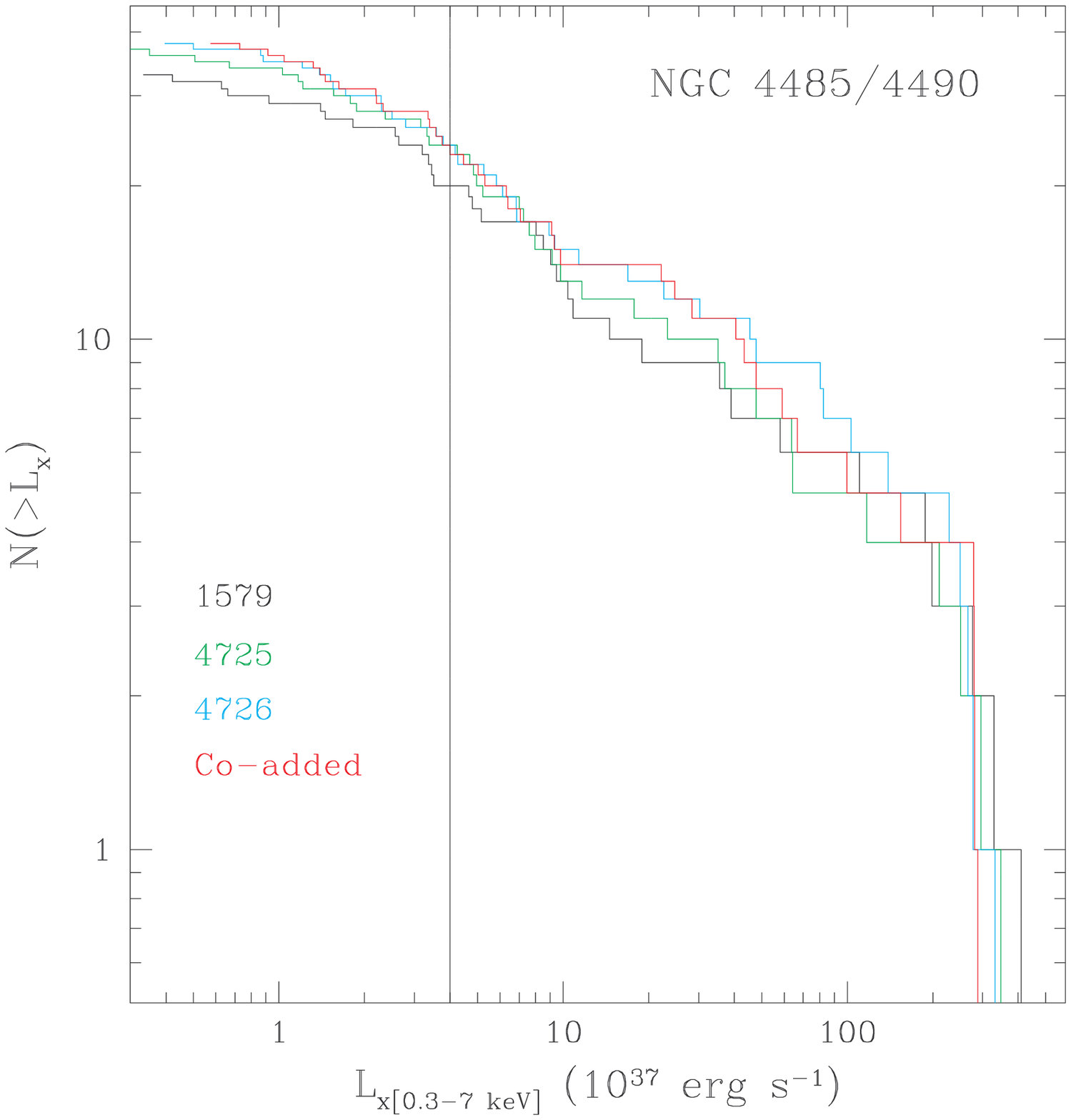}
\caption{Cumulative X-ray luminosity functions for all the {\it Chandra} observations and the co-added exposures of NGC 6946 and NGC 4485/4490. The vertical lines show the adopted completeness limits.}\label{fig:xlf}
\end{figure}

The XLFs for NGC 6946 are shown in Figure~\ref{fig:xlf}. We adopt a completeness limit of $2\times10^{37}\textrm{ ergs s}^{-1}$. Judging from the figure, the XLF of the galaxy does not seem to vary much between observations. This is borne out by the results of the pairwise K-S tests between the five individual observations, which yield significance levels ranging from 0.52 to 0.99 and thus do not warrant rejection of the null hypothesis of a common parent distribution. Interestingly, performing the K-S test between the co-added exposure and each of the observations, we get lower significance levels, ranging from 0.13 to 0.77. The best-fit cumulative slopes of the XLFs are shown in Table~\ref{xlf6946}. They are in all cases $\sim0.6-0.7$ and are in good agreement with each other. We note that \citet{holt2003} derive a slope of $0.68\pm0.03$ for the first {\it Chandra} observation, in agreement with our value of $0.61\pm0.13$.

The XLFs for NGC 4485/90 are shown in Figure~\ref{fig:xlf}. Here we adopt a completeness limit of $4\times10^{37}\textrm{ ergs s}^{-1}$. Again, there does not seem to be significant variation between individual observations. This is supported by pairwise K-S tests between the three observations, which yield significance levels from 0.58 to 0.97. Performing the K-S test between the co-added exposure and individual observations yields higher significance levels here (between 0.93 and 0.98), in contrast to NGC 6946. The best-fit cumulative slopes of the XLFs shown in Table~\ref{xlf4485} agree well with each other and are all $\sim0.4-0.5$. We note that \citet{roberts2002} derive a slope of $0.57\pm0.10$ for the first {\it Chandra} observation, in agreement with our value of $0.45\pm0.10$. There is an indication in Figure~\ref{fig:xlf} of a high-luminosity break or cutoff in the XLFs, which steepen considerably for the $\sim3$ most luminous sources. We therefore also tried fitting a broken power law to the XLFs. This typically led to a best-fit break luminosity of $\sim(2-3)\times10^{39}\textrm{ ergs s}^{-1}$ and a slope for the low-luminosity portion that is $\sim0.05$ lower than for the simple unbroken power law (i.e., $\sim0.4$). The slope of the high-luminosity portion was, however, in all cases very poorly constrained.

We use the results of \cite{bauer2004} to assess whether CBSs are likely to have a significant effect on the XLFs. In NGC 6946 we expect $\sim8$ CBSs among the $\sim25$ sources typically above the completeness limit and within the region covered by all five {\it Chandra} observations. In NGC 4485/90 we estimate that $\sim3$ of the $\sim23$ sources typically above the completeness limit can be expected to be CBSs. Based on these numbers alone, one might expect that CBSs could have a significant effect in the case of NGC 6946 but probably only a small effect for NGC 4485/90. As it turns out, the cumulative number-flux slopes for CBSs derived in \cite{bauer2004} are quite similar to the ones for the XLFs: 0.55 for the $0.5-2$ keV band and 0.56 for the $2-8$ keV band. Contamination of our sample by CBSs is therefore likely to have only a small effect in both cases. The true slopes for NGC 4485/90 are probably slightly flatter than those derived above and slightly steeper for NGC 6946. However, the change in XLF slopes due to CBSs is almost certainly well within the uncertainties quoted in Tables~\ref{xlf6946} and~\ref{xlf4485}.

Our results agree well with previous results for other galaxies. As mentioned in \S~\ref{sec:intro}, XLFs have been seen to be remarkably stable in spite of variability among individual sources, and NGC 6946 and NGC 4485/90 do not seem to be an exception. Single-observation snapshots of their XLFs therefore seem to represent rather well their shape averaged over longer times. Their cumulative XLF slopes fall firmly in the regime of starburst galaxies, whose slopes are typically between 0.4 and 0.8 \citep{fabbiano2006b}---another clear indication of the young stellar populations in these two systems.

\begin{deluxetable}{cc}
\tablewidth{0pt}
\tabletypesize{\scriptsize}
\tablecaption{Power-law Indices of the XLFs of NGC 4485/90\label{xlf4485}}
\tablehead{\colhead{Obs.} & \colhead{$\alpha$\tablenotemark{a}}}
\startdata
1579 & $0.45\pm0.10$\\
4725 & $0.49\pm0.09$\\
4726 & $0.42\pm0.08$\\
Co-added & $0.45\pm0.08$\\
\enddata
\tablenotetext{a}{Best-fit cumulative slope. Uncertainties given are estimated 90\% confidence intervals.}
\end{deluxetable}

\section{Summary and Conclusions}\label{sec:conclusions}

We have presented a study of the X-ray source populations in the spiral galaxy NGC 6946 and the irregular/spiral interacting galaxy pair NGC 4485/4490. We analyzed data from five {\it Chandra} observations of NGC 6946 and from three {\it Chandra} observations of NGC 4485/90. We detect 90 point sources coincident with NGC 6946 (excluding a small circumnuclear region) down to luminosities of a few times $10^{36}\textrm{ ergs s}^{-1}$. We associate five of these with foreground stars. Seven sources are coincident with optically or radio identified SNRs, and four are supernovae from the past 40 years. We detect 38 sources coincident with NGC 4485/90 down to a luminosity of $\sim1\times10^{37}\textrm{ ergs s}^{-1}$.

In NGC 6946, 25 of the 85 sources exhibit long-term (weeks to years) variability in luminosity; 11 are transient candidates. We detect short-term (hours) variability in four sources. Three sources show significant variability in their hardness ratios. We find that the single ULX in NGC 6946, the SNR known as MF16, exhibits long-term variability in flux and color that strongly supports speculations of its high luminosity arising from an XRB within the remnant.

In NGC 4485/90, 15 of the 38 sources show flux variability on timescales of months to years; four are transient candidates. Short-term variability is seen for four sources, and three show variability in their hardness ratios. NGC 4485/90 contains eight ULXs. Two of these have not been identified as ULXs before; this shows the value of long-term monitoring in identifying ultraluminous sources. All but one of the ULXs exhibit long-term variability, and one of them is a transient. One of the ULXs, which has been associated with a SNR based on spectral characteristics and its spatial coincidence with a radio source, shows clear long-term variability pointing to an accreting XRB. We detect short-term variability for one ULX, and two have significant variations in color. This abundant variability among the ULXs clearly points to accretion as the source of luminosity.

Consistent with the widespread variability seen within the source populations, the X-ray colors of the sources indicate that the populations are dominated by XRBs. The source population of NGC 6946 also seems to contain a sizable SNR component, perhaps $\sim15\%-20\%$ of the detected sources. The distribution of colors among the sources in both NGC 6946 and NGC 4485/90 indicates roughly equal fractions of LMXBs and HMXBs. Such a significant presence of HMXBs is consistent with the fact that both systems show high star formation activity.

The shape of the XLFs of NGC 6946 and NGC 4485/90 does not change significantly between observations, and seems to be well represented by single-observation snapshots. The XLF of NGC 6946 (above $2\times10^{37}\textrm{ ergs s}^{-1}$) can be described by a power law with cumulative slope $\sim0.6-0.7$. The XLF of NGC 4485/90 can be described by a power law with cumulative slope $\sim0.4$ between $4\times10^{37}$ and $2\times10^{39}\textrm{ ergs s}^{-1}$. Above $\sim2\times10^{39}\textrm{ ergs s}^{-1}$, the luminosity function drops off rather sharply. The flatness of the slopes (especially for NGC 4485/90) is another testament to the systems' high star formation rate and young stellar population.

\acknowledgements
This work was supported by {\it Chandra} Award AR6-7009X. Use was made of data obtained from the High Energy Astrophysics Science Archive Research Center (HEASARC), provided by NASA's Goddard Space Flight Center. We thank the referee for thorough reading of the paper and constructive comments.

\clearpage
\LongTables
\begin{landscape}
\tabletypesize{\scriptsize}
\def\arraystretch{1.06}
\tabcolsep=3pt
\begin{deluxetable}{ccccccccccccccl}
\tablewidth{0pt}
\tablecolumns{15}
\tablecaption{Basic {\it Chandra} Source Properties for NGC 6946\label{source_prop6946}}
\tablehead{\multicolumn{2}{c}{} & \multicolumn{2}{c}{Position (J2000.0)} & \multicolumn{1}{c}{S/N} & \multicolumn{6}{c}{Photon Flux\tablenotemark{b} ($10^{-7}\textrm{ photons}\textrm{ s}^{-1}\textrm{ cm}^{-2}$)} & \multicolumn{1}{c}{} & \multicolumn{2}{c}{Lum.\tablenotemark{d} ($10^{37}\textrm{ ergs}\textrm{ s}^{-1}$)} & \multicolumn{1}{c}{} \\
\cline{3-4} \cline{6-11} \cline{13-14}
\colhead{No.} & \colhead{Source Designation} & \colhead{R. A.} & \colhead{Decl.} & \colhead{Ratio\tablenotemark{a}} & \colhead{Obs. ID} & \colhead{Obs. ID} & \colhead{Obs. ID} & \colhead{Obs. ID} & \colhead{Obs. ID} & \colhead{Comb.\tablenotemark{c}} & \colhead{} & \colhead{Avg.\tablenotemark{e}} & \colhead{Max.\tablenotemark{f}} & \colhead{Comments\tablenotemark{g}}\\
\colhead{} & \colhead{CXOU J} & \colhead{} & \colhead{} & \colhead{($\sigma$)} & \colhead{1043} & \colhead{4404} & \colhead{4631} & \colhead{4632} & \colhead{4633} & \colhead{} & \colhead{} & \colhead{} & \colhead{}}
\startdata
1	&	203419.0+601007	&	308.57942	&	60.16884	&	7.2	&	$	23.1	\pm	3.3			$	&	$	9.7	\pm	4.1			$	&		\nodata						&		\nodata						&	$	<54.5					$	&	$	18.0	\pm	2.4			$	&&	2.2	&	2.5	&		\\
2	&	203421.4+601017	&	308.58923	&	60.17162	&	7.6	&	$	19.6	\pm	3.5			$	&	$	36.3	\pm	6.6			$	&		\nodata						&		\nodata						&	$	<54.7					$	&	$	24.1	\pm	3.0			$	&&	5.5	&	8.1	&		\\
3	&	203423.4+601039	&	308.59770	&	60.17773	&	4.8	&	$	<4.0					$	&	$	21.3	\pm	4.7			$	&		\nodata						&		\nodata						&	$	29	\pm	12			$	&	$	8.9	\pm	1.7			$	&&	1.1	&	3.4	&	T, N	\\
4	&	203424.7+601038	&	308.60311	&	60.17737	&	7.9	&	$	30.5	\pm	4.3			$	&	$	13.6	\pm	4.2			$	&		\nodata						&		\nodata						&	$	41	\pm	18			$	&	$	25.6	\pm	3.1			$	&&	6.6	&	8.4	&		\\
5	&	203425.0+600906	&	308.60426	&	60.15171	&	13.3	&	$	80.1	\pm	6.1			$	&	$	<8.5				$	&		\nodata						&		\nodata						&	$	19.9	\pm	5.7			$	&	$	47.7	\pm	3.5			$	&&	7.1	&	12	&	T	\\
6	&	203426.1+600909	&	308.60889	&	60.15262	&	32.1	&	$	343	\pm	12			$	&	$	155	\pm	12			$	&		\nodata						&		\nodata						&	$	127	\pm	16			$	&	$	261.4	\pm	8.0			$	&&	32	&	44	&		\\
7	&	203426.2+600914	&	308.60940	&	60.15414	&	7.1	&	$	25.9	\pm	4.1			$	&	$	17.1	\pm	5.4			$	&		\nodata						&		\nodata						&	$	24	^{+	15	}_{-	10	}$	&	$	23.2	\pm	3.1			$	&&	0.9	&	1.0	&	SSS	\\
8	&	203429.1+601051	&	308.62137	&	60.18085	&	6.5	&	$	11.5	\pm	2.4			$	&	$	28.4	\pm	5.5			$	&		\nodata						&		\nodata						&	$	<22.0					$	&	$	15.2	\pm	2.2			$	&&	2.2	&	4.8	&	T	\\
\enddata
\tablenotetext{a}{Photometric signal-to-noise ratio based on all observations, as computed by ACIS Extract.}
\tablenotetext{b}{Background-subtracted photon fluxes in the full $0.3-7$ keV band. For sources with fewer than five net counts in an observation an upper limit with 3 $\sigma$ Gaussian confidence is given.}
\tablenotetext{c}{Photon flux in the co-added exposure.}
\tablenotetext{d}{X-ray luminosity in the $0.3-7$ keV band.}
\tablenotetext{e}{\phantom{}Luminosity calculated from a co-added exposure spanning all the observations.}
\tablenotetext{f}{\phantom{}Luminosity calculated from the observation in which the source is brightest.}
\tablenotetext{g}{(T) transient candidate; (N) source not detected by \cite{holt2003}; (SSS) supersoft source candidate; (S) source is most likely a foreground star; (LD) source coincides with a supernova remnant in the list of \citet{lacey2001}; (MF) source coincides with a supernova remnant in the list of \citet{matonick1997}.}
\end{deluxetable}
\clearpage
\end{landscape}

\clearpage
\LongTables
\begin{landscape}
\tabletypesize{\scriptsize}
\def\arraystretch{1.06}
\begin{deluxetable}{ccccccccccccl}
\tablewidth{0pt}
\tablecolumns{13}
\tablecaption{Basic {\it Chandra} Source Properties for NGC 4485/4490\label{source_prop4485}}
\tablehead{\multicolumn{2}{c}{} & \multicolumn{2}{c}{Position (J2000.0)} & \multicolumn{1}{c}{S/N} & \multicolumn{4}{c}{Photon Flux\tablenotemark{b} ($10^{-7}\textrm{ photons}\textrm{ s}^{-1}\textrm{ cm}^{-2}$)} & \multicolumn{1}{c}{} & \multicolumn{2}{c}{Lum.\tablenotemark{d} ($10^{37}\textrm{ ergs s}^{-1}$)} & \multicolumn{1}{c}{}\\
\cline{3-4} \cline{6-9} \cline{11-12}
\colhead{No.} & \colhead{Source Designation} & \colhead{R. A.} & \colhead{Decl.} & \colhead{Ratio\tablenotemark{a}} & \colhead{Obs. ID} & \colhead{Obs. ID} & \colhead{Obs. ID} & \colhead{Comb.\tablenotemark{c}}  & \colhead{} & \colhead{Avg.\tablenotemark{e}} & \colhead{Max.\tablenotemark{f}} & \colhead{Comments\tablenotemark{g}}\\
\colhead{} & \colhead{CXOU J} & \colhead{} & \colhead{} & \colhead{($\sigma$)} & \colhead{1579} & \colhead{4725} & \colhead{4726} & \colhead{} & \colhead{} & \colhead{} & \colhead{}}
\startdata
1	&	123025.2+413924	&	187.60506	&	41.65691	&	15.6	&	$	73	\pm	10			$	&	$	70.5	\pm	7.5			$	&	$	77.5	\pm	7.5			$	&	$	73.9	\pm	4.6			$	&&	22	&	23	&		\\
2	&	123026.7+413821	&	187.61165	&	41.63942	&	5.2	&	$	<9.4					$	&	$	15.0	\pm	3.7			$	&	$	10.2	\pm	2.8			$	&	$	9.8	\pm	1.8			$	&&	2.2	&	3.3	&	T, N	\\
3	&	123027.1+413814	&	187.61330	&	41.63724	&	8.7	&	$	25.1	\pm	6.5			$	&	$	27.0	\pm	4.8			$	&	$	25.1	\pm	4.5			$	&	$	25.7	\pm	2.8			$	&&	9.3	&	10	&		\\
4	&	123027.6+413941	&	187.61528	&	41.66161	&	4.1	&	$	<17.9					$	&	$	2.8	^{+	1.9	}_{-	1.3	}$	&	$	12.4	\pm	3.3			$	&	$	6.7	\pm	1.5			$	&&	1.5	&	2.8	&	N	\\
5	&	123028.2+413958	&	187.61779	&	41.66622	&	4.2	&	$	18.4	\pm	5.3			$	&	$	6.6	\pm	2.9			$	&	$	<12.7					$	&	$	7.4	\pm	1.6			$	&&	1.3	&	3.2	&		\\
6	&	123028.5+413926	&	187.61913	&	41.65731	&	5.6	&	$	19.7	\pm	5.9			$	&	$	11.5	\pm	3.3			$	&	$	10.5	\pm	3.3			$	&	$	12.8	\pm	2.1			$	&&	5.3	&	8.0	&		\\
7	&	123028.7+413757	&	187.61969	&	41.63254	&	9.5	&	$	12.1	\pm	4.3			$	&	$	20.1	\pm	4.4			$	&	$	58.9	\pm	7.3			$	&	$	33.7	\pm	3.4			$	&&	9.8	&	17	&		\\
8	&	123029.0+414046	&	187.62122	&	41.67961	&	6.9	&	$	19.4	\pm	5.7			$	&	$	20.8	\pm	4.3			$	&	$	13.1	\pm	3.4			$	&	$	17.4	\pm	2.4			$	&&	3.7	&	4.8	&		\\
\enddata
\tablenotetext{a}{Photometric signal-to-noise ratio based on all observations, as computed by ACIS Extract.}
\tablenotetext{b}{Background-subtracted photon fluxes in the full $0.3-7$ keV band. For sources with fewer than five net counts in an observation an upper limit with 3 $\sigma$ Gaussian confidence is given.}
\tablenotetext{c}{Photon flux in the co-added exposure.}
\tablenotetext{d}{Observed X-ray luminosity in the $0.3-7$ keV band.}
\tablenotetext{e}{\phantom{}Luminosity calculated from a co-added exposure spanning all the observations.}
\tablenotetext{f}{\phantom{}Luminosity calculated from the observation in which the source is brightest.}
\tablenotetext{g}{(T) transient candidate; (N) source not detected by \cite{roberts2002}; (SSS) supersoft source candidate.}
\end{deluxetable}
\clearpage
\end{landscape}


\clearpage
\tabletypesize{\scriptsize}
\tabcolsep=3pt
\def\arraystretch{1.06}
\begin{deluxetable}{cccccccccccccl}
\tablewidth{0pt}
\tablecolumns{14}
\tablecaption{{\it Chandra} Variability Properties for NGC 6946 Sources\label{var_prop6946}}
\tablehead{\multicolumn{2}{c}{} & \multicolumn{5}{c}{K-S Test for Short-Term Flux Variability\tablenotemark{a}} & \multicolumn{1}{c}{} & \multicolumn{5}{c}{Long-Term Variability} & \multicolumn{1}{c}{}\\
\cline{3-7} \cline{9-13}
\colhead{No.} & \colhead{Source Designation} & \colhead{Obs. ID} & \colhead{Obs. ID} & \colhead{Obs. ID}  & \colhead{Obs. ID} & \colhead{Obs. ID} & \colhead{} & \multicolumn{2}{c}{Flux} & \colhead{} & \multicolumn{2}{c}{Hardness Ratios} & \colhead{Comments\tablenotemark{f}}\\
\cline{9-10} \cline{12-13}
\colhead{} & \colhead{CXOU J} & \colhead{1043} & \colhead{4404} & \colhead{4631} & \colhead{4632} & \colhead{4633} & \colhead{} & \colhead{$S_\mathrm{flux}$\tablenotemark{b}} & \colhead{$R$\tablenotemark{c}} & \colhead{} & \colhead{$S_\mathrm{HR1}$\tablenotemark{d}} & \colhead{$S_\mathrm{HR2}$\tablenotemark{e}} & \colhead{}}
\startdata
1	&	203419.0+601007	&	0.12	&	0.94	&	\nodata	&	\nodata	&	0.83	&&	2.2	&	2.4	&&	0.5	&	0.8	&		\\
2	&	203421.4+601017	&	0.96	&	0.98	&	\nodata	&	\nodata	&	0.39	&&	2.0	&	3.9	&&	0.6	&	0.2	&		\\
3	&	203423.4+601039	&	\nodata	&	0.66	&	\nodata	&	\nodata	&	0.13	&&	4.0	&	110	&&	0.7	&	0.5	&	l	\\
4	&	203424.7+601038	&	0.65	&	0.68	&	\nodata	&	\nodata	&	0.29	&&	2.5	&	3.0	&&	0.9	&	1.4	&		\\
5	&	203425.0+600906	&	0.30	&	\nodata	&	\nodata	&	\nodata	&	0.15	&&	11.9	&	73.0	&&	0.8	&	0.5	&	l	\\
6	&	203426.1+600909	&	0.48	&	0.69	&	\nodata	&	\nodata	&	0.77	&&	10.9	&	2.7	&&	0.6	&	1.6	&	l	\\
7	&	203426.2+600914	&	0.58	&	0.47	&	\nodata	&	\nodata	&	0.49	&&	1.2	&	1.5	&&	0.4	&	0.3	&		\\
8	&	203429.1+601051	&	0.33	&	0.85	&	\nodata	&	\nodata	&	\nodata	&&	3.2	&	$>1.5$	&&	0.5	&	0.9	&	l	\\
\enddata
\tablenotetext{a}{Significance level of a one-sided Kolmogorov-Smirnov test statistic with respect to a uniform count-rate model.}
\tablenotetext{b}{$S_\mathrm{flux}=\max_{i,j}|F_i-F_j|/\sqrt{\sigma_{F_i}^2+\sigma_{F_j}^2}$}
\tablenotetext{c}{$R=F_\mathrm{max}/F_\mathrm{min}$. For sources with no detected flux in at least one of the observations a lower limit with 3 $\sigma$ Gaussian confidence is given.}
\tablenotetext{d}{$S_\mathrm{HR1}=\max_{i,j}{|\textrm{HR1}_i-\textrm{HR1}_j|/\sqrt{\sigma_{\mathrm{HR1},i}^2+\sigma_{\mathrm{HR1},j}^2}}$}
\tablenotetext{e}{$S_\mathrm{HR2}=\max_{i,j}{|\textrm{HR2}_i-\textrm{HR2}_j|/\sqrt{\sigma_{\mathrm{HR2},i}^2+\sigma_{\mathrm{HR2},j}^2}}$}
\tablenotetext{f}{Short-term flux variability is denoted by ``s,'' long-term flux variability by ``l,'' and spectral variability by ``sp.'' Foreground stars are indicated by ``star.''}
\end{deluxetable}

\clearpage

\tabletypesize{\scriptsize}
\def\arraystretch{1.06}
\begin{deluxetable}{cccccccccccl}
\tablewidth{0pt}
\tablecolumns{12}
\tablecaption{{\it Chandra} Variability Properties for NGC 4485/4490 Sources\label{var_prop4485}}
\tablehead{\multicolumn{2}{c}{} & \multicolumn{3}{c}{K-S Test for Short-Term Flux Variability\tablenotemark{a}} & \multicolumn{1}{c}{} & \multicolumn{5}{c}{Long-Term Variability} & \multicolumn{1}{c}{}\\
\cline{3-5} \cline{7-11}
\colhead{No.} & \colhead{Source Designation} & \colhead{Obs. ID} & \colhead{Obs. ID} & \colhead{Obs. ID}  & \colhead{} & \multicolumn{2}{c}{Flux} & \colhead{} & \multicolumn{2}{c}{Hardness Ratios} & \colhead{Comments\tablenotemark{f}}\\
\cline{7-8} \cline{10-11}
\colhead{} & \colhead{CXOU J} & \colhead{1579} & \colhead{4725} & \colhead{4726} & \colhead{} & \colhead{$S_\mathrm{flux}$\tablenotemark{b}} & \colhead{$R$\tablenotemark{c}} & \colhead{} & \colhead{$S_\mathrm{HR1}$\tablenotemark{d}} & \colhead{$S_\mathrm{HR2}$\tablenotemark{e}} & \colhead{}}
\startdata
1	&	123025.2+413924	&	0.35	&	0.50	&	0.61	&&	0.6	&	1.1	&&	1.1	&	1.8	&		\\
2	&	123026.7+413821	&	\nodata	&	0.25	&	1.00	&&	3.1	&	$>1.6$	&&	1.2	&	1.6	&	l	\\
3	&	123027.1+413814	&	0.84	&	0.72	&	0.35	&&	0.3	&	1.1	&&	0.5	&	1.3	&		\\
4	&	123027.6+413941	&	\nodata	&	0.35	&	0.77	&&	2.3	&	4.5	&&	0.8	&	0.8	&		\\
5	&	123028.2+413958	&	0.10	&	0.73	&	0.42	&&	2.3	&	5.4	&&	1.3	&	0.5	&		\\
6	&	123028.5+413926	&	0.82	&	0.05	&	0.94	&&	1.2	&	1.9	&&	1.3	&	0.8	&		\\
7	&	123028.7+413757	&	0.87	&	0.84	&	0.90	&&	5.1	&	4.9	&&	1.3	&	2.0	&	l	\\
8	&	123029.0+414046	&	0.22	&	0.62	&	0.57	&&	1.3	&	1.6	&&	0.5	&	0.3	&		\\
\enddata
\tablenotetext{a}{Significance level of a one-sided Kolmogorov-Smirnov test statistic with respect to a uniform count-rate model.}
\tablenotetext{b}{$S_\mathrm{flux}=\max_{i,j}|F_i-F_j|/\sqrt{\sigma_{F_i}^2+\sigma_{F_j}^2}$}
\tablenotetext{c}{$R=F_\mathrm{max}/F_\mathrm{min}$. For sources with no detected flux in at least one of the observations a lower limit with 3 $\sigma$ Gaussian confidence is given.}
\tablenotetext{d}{$S_\mathrm{HR1}=\max_{i,j}{|\textrm{HR1}_i-\textrm{HR1}_j|/\sqrt{\sigma_{\mathrm{HR1},i}^2+\sigma_{\mathrm{HR1},j}^2}}$}
\tablenotetext{e}{$S_\mathrm{HR2}=\max_{i,j}{|\textrm{HR2}_i-\textrm{HR2}_j|/\sqrt{\sigma_{\mathrm{HR2},i}^2+\sigma_{\mathrm{HR2},j}^2}}$}
\tablenotetext{f}{Short-term flux variability is denoted by ``s,'' long-term flux variability by ``l,'' and spectral variability by ``sp.''}
\end{deluxetable}

\clearpage

\tabletypesize{\scriptsize}
\def\arraystretch{1.06}
\begin{deluxetable}{cccccccccc}
\tablewidth{0pt}
\tablecolumns{10}
\tablecaption{{\it Chandra} Spectral Properties for NGC 6946 Sources\label{spec_prop6946a}}
\tablehead{\multicolumn{2}{c}{} & \multicolumn{8}{c}{Hardness Ratios}\\
\cline{3-10}
\colhead{No.} & \colhead{Source Designation} & \multicolumn{2}{c}{Obs. ID 1043} & \colhead{} &\multicolumn{2}{c}{Obs. ID 4404} & \colhead{} &\multicolumn{2}{c}{Obs. ID 4631}\\
\cline{3-4} \cline{6-7} \cline{9-10}
\colhead{} & \colhead{CXOU J} & \colhead{HR1\tablenotemark{a}} & \colhead{HR2\tablenotemark{b}} & \colhead{} & \colhead{HR1} & \colhead{HR2} & \colhead{} & \colhead{HR1} & \colhead{HR2}}
\startdata
1	&	203419.0+601007	&	$	0.09	\pm	0.10	$	&	$	-0.07	\pm	0.10	$	&&	$	-0.12	\pm	0.37	$	&	$	0.10	\pm	0.35	$	&&		\nodata				&		\nodata					\\
2	&	203421.4+601017	&	$	0.04	\pm	0.08	$	&	$	0.85	\pm	0.25	$	&&	$	-0.02	\pm	0.09	$	&	$	0.92	\pm	0.26	$	&&		\nodata				&		\nodata					\\
3	&	203423.4+601039	&			\nodata			&			\nodata			&&	$	0.20	\pm	0.16	$	&	$	-0.07	\pm	0.17	$	&&		\nodata				&		\nodata					\\
4	&	203424.7+601038	&	$	0.09	\pm	0.05	$	&	$	0.84	\pm	0.19	$	&&	$	0.04	\pm	0.09	$	&	$	1.25	\pm	0.54	$	&&		\nodata				&		\nodata					\\
5	&	203425.0+600906	&	$	0.03	\pm	0.05	$	&	$	0.15	\pm	0.06	$	&&	$	0.2	\pm	1.0	$	&	$	1.6	\pm	3.0	$	&&		\nodata				&		\nodata					\\
6	&	203426.1+600909	&	$	0.05	\pm	0.02	$	&	$	0.06	\pm	0.03	$	&&	$	0.08	\pm	0.05	$	&	$	-0.03	\pm	0.05	$	&&		\nodata				&		\nodata					\\
7	&	203426.2+600914	&	$	-1.10	\pm	0.25	$	&	$	0.00	\pm	0.03	$	&&	$	-1.25	\pm	0.55	$	&	$	-0.01	\pm	0.09	$	&&		\nodata				&		\nodata					\\
8	&	203429.1+601051	&	$	0.13	\pm	0.15	$	&	$	0.01	\pm	0.17	$	&&	$	0.03	\pm	0.13	$	&	$	0.22	\pm	0.17	$	&&		\nodata				&		\nodata					\\
\enddata
\tablenotetext{a}{Soft color: $\mathrm{HR1}=(\mathrm{M}-\mathrm{S})/\mathrm{T}$.}
\tablenotetext{b}{Hard color: $\mathrm{HR2}=(\mathrm{H}-\mathrm{M})/\mathrm{T}$.}
\end{deluxetable}


\tabletypesize{\scriptsize}
\def\arraystretch{1.06}
\begin{deluxetable}{cccccccccc}
\tablewidth{0pt}
\tablecolumns{10}
\tablecaption{{\it Chandra} Spectral Properties for NGC 6946 Sources\label{spec_prop6946b}}
\tablehead{\multicolumn{2}{c}{} & \multicolumn{8}{c}{Hardness Ratios}\\
\cline{3-10}
\colhead{No.} & \colhead{Source Designation} & \multicolumn{2}{c}{Obs. ID 4632} & \colhead{} &\multicolumn{2}{c}{Obs. ID 4633} & \colhead{} &\multicolumn{2}{c}{Combined\tablenotemark{c}}\\
\cline{3-4} \cline{6-7} \cline{9-10}
\colhead{} & \colhead{CXOU J} & \colhead{HR1\tablenotemark{a}} & \colhead{HR2\tablenotemark{b}} & \colhead{} & \colhead{HR1} & \colhead{HR2} & \colhead{} & \colhead{HR1} & \colhead{HR2}}
\startdata
1	&	203419.0+601007	&		\nodata				&		\nodata				&&	$	0.6	\pm	1.4	$	&	$	-0.63	\pm	0.82	$	&&	$	0.08	\pm	0.10	$	&	$	-0.08	\pm	0.10	$		\\
2	&	203421.4+601017	&		\nodata				&		\nodata				&&	$	1.0	\pm	1.8	$	&	$	0.92	\pm	1.65	$	&&	$	0.02	\pm	0.05	$	&	$	0.88	\pm	0.17	$		\\
3	&	203423.4+601039	&		\nodata				&		\nodata				&&	$	-0.27	\pm	0.68	$	&	$	-0.26	\pm	0.33	$	&&	$	0.09	\pm	0.15	$	&	$	-0.12	\pm	0.14	$		\\
4	&	203424.7+601038	&		\nodata				&		\nodata				&&	$	0.39	\pm	0.38	$	&	$	0.32	\pm	0.43	$	&&	$	0.11	\pm	0.04	$	&	$	0.87	\pm	0.17	$		\\
5	&	203425.0+600906	&		\nodata				&		\nodata				&&	$	0.20	\pm	0.21	$	&	$	0.22	\pm	0.26	$	&&	$	0.03	\pm	0.05	$	&	$	0.16	\pm	0.06	$		\\
6	&	203426.1+600909	&		\nodata				&		\nodata				&&	$	0.01	\pm	0.10	$	&	$	-0.04	\pm	0.09	$	&&	$	0.05	\pm	0.02	$	&	$	0.04	\pm	0.02	$		\\
7	&	203426.2+600914	&		\nodata				&		\nodata				&&	$	-1.5	\pm	1.1	$	&	$	-0.07	\pm	0.20	$	&&	$	-1.18	\pm	0.22	$	&	$	-0.01	\pm	0.02	$		\\
8	&	203429.1+601051	&		\nodata				&		\nodata				&&		\nodata				&		\nodata				&&	$	0.08	\pm	0.10	$	&	$	0.13	\pm	0.12	$		\\
\enddata
\tablenotetext{a}{Soft color: $\mathrm{HR1}=(\mathrm{M}-\mathrm{S})/\mathrm{T}$.}
\tablenotetext{b}{Hard color: $\mathrm{HR2}=(\mathrm{H}-\mathrm{M})/\mathrm{T}$.}
\tablenotetext{c}{Hardness ratios calculated from the co-added exposure.}
\end{deluxetable}

\clearpage
\tabletypesize{\scriptsize}
\tabcolsep=3pt
\def\arraystretch{1.06}
\begin{deluxetable}{ccccccccccccc}
\tablewidth{0pt}
\tablecolumns{13}
\tablecaption{{\it Chandra} Spectral Properties for NGC 4485/4490 Sources\label{spec_prop4485}}
\tablehead{\multicolumn{2}{c}{} & \multicolumn{11}{c}{Hardness Ratios}\\
\cline{3-13}
\colhead{No.} & \colhead{Source Designation} & \multicolumn{2}{c}{Obs. ID 1579} & \colhead{} &\multicolumn{2}{c}{Obs. ID 4725} & \colhead{} &\multicolumn{2}{c}{Obs. ID 4726} & \colhead{} &\multicolumn{2}{c}{Combined\tablenotemark{c}}\\
\cline{3-4} \cline{6-7} \cline{9-10} \cline{12-13}
\colhead{} & \colhead{CXOU J} & \colhead{HR1\tablenotemark{a}} & \colhead{HR2\tablenotemark{b}} & \colhead{} & \colhead{HR1} & \colhead{HR2} & \colhead{} & \colhead{HR1} & \colhead{HR2} & \colhead{} & \colhead{HR1} & \colhead{HR2}}
\startdata
1	&	123025.2+413924	&	$	0.13	\pm	0.08	$	&	$	0.08	\pm	0.11	$	&&	$	0.01	\pm	0.07	$	&	$	0.35	\pm	0.10	$	&&	$	0.12	\pm	0.06	$	&	$	0.25	\pm	0.09	$	&&	$	0.08	\pm	0.04	$	&	$	0.25	\pm	0.06	$		\\
2	&	123026.7+413821	&		\nodata				&		\nodata				&&	$	-0.15	\pm	0.22	$	&	$	-0.05	\pm	0.17	$	&&	$	0.22	\pm	0.20	$	&	$	-0.04	\pm	0.22	$	&&	$	0.03	\pm	0.13	$	&	$	-0.04	\pm	0.13	$		\\
3	&	123027.1+413814	&	$	0.15	\pm	0.10	$	&	$	0.74	\pm	0.34	$	&&	$	0.09	\pm	0.12	$	&	$	0.24	\pm	0.16	$	&&	$	0.08	\pm	0.10	$	&	$	0.52	\pm	0.20	$	&&	$	0.10	\pm	0.06	$	&	$	0.45	\pm	0.11	$		\\
4	&	123027.6+413941	&	$	-0.12	\pm	0.69	$	&	$	0.24	\pm	0.82	$	&&	$	0.36	\pm	0.49	$	&	$	-0.31	\pm	0.46	$	&&	$	-0.05	\pm	0.21	$	&	$	0.09	\pm	0.21	$	&&	$	0.03	\pm	0.17	$	&	$	0.02	\pm	0.17	$		\\
5	&	123028.2+413958	&	$	-0.06	\pm	0.23	$	&	$	-0.24	\pm	0.19	$	&&	$	-0.51	\pm	0.53	$	&	$	-0.10	\pm	0.29	$	&&	$	0.49	\pm	0.55	$	&	$	0.10	\pm	0.65	$	&&	$	-0.14	\pm	0.19	$	&	$	-0.15	\pm	0.14	$		\\
6	&	123028.5+413926	&	$	0.13	\pm	0.16	$	&	$	0.55	\pm	0.35	$	&&	$	0.33	\pm	0.17	$	&	$	0.32	\pm	0.29	$	&&	$	-0.02	\pm	0.20	$	&	$	0.70	\pm	0.40	$	&&	$	0.16	\pm	0.09	$	&	$	0.51	\pm	0.18	$		\\
7	&	123028.7+413757	&	$	-0.20	\pm	0.30	$	&	$	-0.15	\pm	0.23	$	&&	$	-0.07	\pm	0.15	$	&	$	0.41	\pm	0.22	$	&&	$	-0.31	\pm	0.11	$	&	$	0.36	\pm	0.11	$	&&	$	-0.23	\pm	0.08	$	&	$	0.33	\pm	0.09	$		\\
8	&	123029.0+414046	&	$	-0.11	\pm	0.22	$	&	$	0.09	\pm	0.24	$	&&	$	-0.07	\pm	0.17	$	&	$	0.03	\pm	0.16	$	&&	$	-0.24	\pm	0.25	$	&	$	-0.01	\pm	0.19	$	&&	$	-0.13	\pm	0.11	$	&	$	0.03	\pm	0.10	$		\\
\enddata
\tablenotetext{a}{Soft color: $\mathrm{HR1}=(\mathrm{M}-\mathrm{S})/\mathrm{T}$.}
\tablenotetext{b}{Hard color: $\mathrm{HR2}=(\mathrm{H}-\mathrm{M})/\mathrm{T}$.}
\tablenotetext{c}{Hardness ratios calculated from the co-added exposure.}
\end{deluxetable}

\end{document}